\documentclass[
reprint,
superscriptaddress,
 amsmath,amssymb,
 aps,
 prx,
]{revtex4-2}

\usepackage{graphicx}
\usepackage{dcolumn}
\usepackage{bm}
\usepackage[dvipsnames]{xcolor}
\usepackage{hyperref}
\hypersetup{
	colorlinks,
	linkcolor={red!50!black},
	citecolor={red!70!black},
	urlcolor={red!80!black}
}
\usepackage[capitalize]{cleveref}
\usepackage{booktabs}
\usepackage{xspace}
\usepackage[version=4]{mhchem}
\usepackage{newtxtext,newtxmath}

\newcommand{\kmesh}{$k$-mesh\xspace}
\newcommand{\kmeshes}{$k$-meshes\xspace}

\newcommand{\gwhf}{G$_0$W$_0$@HF\xspace}

\newcommand{\mesh}[1]{$N_k ={#1}^3$\xspace}
\newcommand{\meshto}[2]{$N_k ={#1}^3$--${#2}^3$\xspace}

\usepackage{multirow}
\usepackage{siunitx}
\usepackage{braket}

\usepackage{lipsum}

\begin{document}

\preprint{APS/123-QED}

\title{Reaching the thermodynamic limit of periodic CCSD cohesive energies and band gaps}

\author{Shuhang Li}
 \affiliation{Department of Chemistry and Cherry Emerson Center for Scientific Computation, Emory University, Atlanta, Georgia 30322, United States}
\author{Huanchen Zhai}
 \affiliation{Initiative for Computational Catalysis, Flatiron Institute, New York, New York 10010, United States}
\author{Francesco A. Evangelista}
 \affiliation{Department of Chemistry and Cherry Emerson Center for Scientific Computation, Emory University, Atlanta, Georgia 30322, United States}
\author{Timothy C. Berkelbach}
 \email{tberkelbach@flatironinstitute.org}
 \affiliation{Initiative for Computational Catalysis, Flatiron Institute, New York, New York 10010, United States}
 \affiliation{Department of Chemistry, Columbia University, New York, New York, 10027, United States}

\date{\today}

\begin{abstract}
The high computational cost of periodic coupled-cluster theory has limited the density of Brillouin zone sampling, yielding finite-size errors that need to be removed by extrapolation.
Here we report the development and application of a distributed-memory software implementation of periodic coupled-cluster theory with single and double excitations (CCSD) that runs efficiently on up to 12 nodes with 96 cores each.
This new implementation allows ground-state and excited-state calculations in which the Brillouin zone is sampled with up to $6^3=216$ $k$-points, allowing us to reliably extrapolate to the thermodynamic limit.
For eight simple semiconductors and insulators, we report the cohesive energy and band gap, which are converged to 0.1~eV, providing definitive benchmark numbers for the CCSD level of theory.
Compared to experimental values, average errors for the cohesive energy are 0.1--0.2~eV (typically an underestimate), and average errors for the band gap are about 0.4~eV (typically an overestimate).
\end{abstract}

\maketitle

\section{Introduction}
\label{sec:intro}

Periodic coupled-cluster theory with single and double excitations (CCSD)~\cite{hirata.2001.10.1016/S0009-26140100897-1, hirata.2004.10.1063/1.1637577, katagiri.2005.10.1063/1.1929731, gruneis.2011.10.1021/ct200263g, mcclain.2017.10.1021/acs.jctc.7b00049} is
increasingly being used to calculate the electronic structure of solids.
It is a nonperturbative, size-extensive many-body theory that can be systematically improved by including higher-order excitations, which makes it a promising method towards achieving high-accuracy predictions and benchmarking more affordable approaches.
Example applications include ground-state properties, such as lattice constants, bulk moduli, and cohesive
energies~\cite{nolan.2009.10.1103/PhysRevB.80.165109, booth.2013.10.1038/nature11770, mcclain.2017.10.1021/acs.jctc.7b00049, gruber.2018.10.1103/PhysRevX.8.021043, neufeld.2022.10.1021/acs.jpclett.2c01828, neufeld.2023.10.1103/PhysRevLett.131.186402, ye.2024.10.1021/acs.jctc.4c00936}, and excited-state properties studied with the
equation-of-motion (EOM) formalism, such as band gaps and neutral excitation
energies~\cite{katagiri.2005.10.1063/1.1929731, lewis.2020.10.1021/acs.jpclett.0c00031, wang.2020.10.1021/acs.jctc.0c00101, gao.2020.10.1103/PhysRevB.101.165138, wang.2021.10.1021/acs.jctc.1c00692, gallo.2021.10.1063/5.0035425,  vo.2024.10.1063/5.0187856, moerman.2025.10.1021/acs.jctc.4c01451, moerman.2025.10.1103/PhysRevB.111.L121202, vo.2025.10.48550/arXiv.2508.00168}.

In electronic structure calculations of solids with periodic boundary conditions, the physically relevant thermodynamic limit (TDL) is reached by increasing the number of $k$-points sampled from the Brillouin zone (or equivalently the number of unit cells within a supercell). The relatively low cost of mean-field theories, including density functional theory (DFT) and Hartree-Fock (HF) theory, allows for a dense sampling of the Brillouin zone, making it straightforward to eliminate finite-size errors. In contrast, the high cost of CCSD---whose storage scales as $O(n^4N_k^3)$ and whose execution time scales as $O(n^6 N_k^4)$, where $n$ is the number of atoms in the unit cell and $N_k$ is the number of $k$-points---has limited the $k$-point sampling to meshes no larger than $N_k = 4^3$ for three-dimensional solids. 

Although the asymptotic, large-$N_k$ scaling of the finite-size error is understood for CCSD ground-state energies~\cite{mcclain.2017.10.1021/acs.jctc.7b00049, mihm.2023.10.1021/acs.jctc.2c00737, xing.2024.10.1103/PhysRevX.14.011059} and band gaps~\cite{mcclain.2017.10.1021/acs.jctc.7b00049, gao.2020.10.1103/PhysRevB.101.165138, vo.2024.10.1063/5.0187856, moerman.2025.10.1021/acs.jctc.4c01451, moerman.2025.10.1103/PhysRevB.111.L121202}, the predictions from small, accessible $k$-point meshes are not guaranteed to be in this limit, precluding reliable extrapolation to the TDL. Specifically, ground-state energies are known to approach the TDL as
\begin{equation}
    E (N_k) = E (\infty) + A N_k^{-1} + \ldots
\end{equation}
whereas band gaps are known to approach the TDL as
\begin{equation}
    E_{\mathrm{gap}} (N_k) = E_{\mathrm{gap}} (\infty) + A'N_k^{-1/3} + \ldots
\end{equation}
This slower convergence of band gaps has made them especially hard to predict reliably with access to only small $k$-point meshes.
Therefore, to precisely assess the accuracy of CCSD for solids, it is necessary to eliminate finite-size errors along with compounding basis set and pseudopotential errors.

In this work, we report the results of a new and optimized parallel implementation of ground-state CCSD and EOM-CCSD.
To the best of our knowledge, these are the largest periodic CCSD calculations based on canonical (delocalized) orbitals to date, using sufficiently large $k$-point sampling for reliable TDL extrapolation.
We use our new implementation to report what we believe are converged CCSD cohesive energies and band gaps of eight simple semiconductors and insulators, and we use our results to evaluate previously employed extrapolation schemes for smaller $k$-point meshes.

\section{Results}
\label{sec:results}
\subsection{Cohesive energies}
\label{sec:cohesive}

By extrapolating results obtained with large \kmeshes and basis sets, we are able to reliably produce converged HF, second-order M{\o}ller-Plesset perturbation theory (MP2), and CCSD cohesive energies for all solids studied in this work.
In \cref{tab:cohesive}, we compare our predicted cohesive energies with experimental values corrected for zero-point energy effects~\cite{zhang.2018.10.1088/1367-2630/aac7f0,schimka.2011.10.1063/1.3524336}.
HF significantly underestimates the cohesive energies, both with and without GTH pseudopotentials (PPs) optimized for Hartree-Fock theory (GTH-HF-rev),\cite{goedecker.1996.10.1103/PhysRevB.54.1703a,hartwigsen.1998.10.1103/PhysRevB.58.3641} while MP2 and CCSD provide substantial improvements.
We find that the PP error is similar at all three levels of theory, indicating that its dominant contribution is determined at the HF level, in agreement with recent works~\cite{ye.2024.10.1021/acs.jctc.4c00936, zhang.2026.10.48550/arXiv.2602.16679}.
Nonetheless, this PP error is usually quite small for most of these solids: all-electron frozen-core (AE-FC) calculations increase the cohesive energy by less than 0.05~eV for most solids and levels of theory.
The most notable outliers are BN and C, for which the cohesive energy is increased by 0.1--0.2~eV at all levels of theory (consistent with a previous finding for C~\cite{ye.2024.10.1021/acs.jctc.4c00936}). 

Because of these relatively small errors, the qualitative trends are the same with and without PPs when comparing to experimental values.
MP2 tends to overpredict the cohesive energy (by 0.24--0.30~eV on average) and CCSD tends to underpredict the cohesive energy (by 0.14--0.19~eV on average).
While the CCSD accuracy is slightly degraded by PPs, the MP2 accuracy is improved, indicating some fortuitous cancellation of error. 
Our extrapolated results show that the accuracy of CCSD is comparable to or better than that of commonly used density functionals, including PBE, PBEsol, and SCAN~\cite{tran.2016.10.1063/1.4948636}, and separate work has demonstrated that CCSD(T) is even more accurate than CCSD for simple semiconductors and insulators~\cite{booth.2013.10.1038/nature11770, gruber.2018.10.1103/PhysRevX.8.021043, ye.2024.10.1021/acs.jctc.4c00936}.

\begin{table}[t]
\centering
\caption{Predicted cohesive energies (in eV/atom) obtained with GTH-HF pseudopotentials (PP) or the all-electron (AE) and frozen-core (FC) approximation at each level of theory, along with the mean absolute error (MAE) and mean absolute relative error (MARE), evaluated at the experimental lattice constants. Experimental values corrected for zero-point energy are from Ref.~\citenum{zhang.2018.10.1088/1367-2630/aac7f0}, except for the \ce{LiH} value, which is taken from Ref.~\citenum{schimka.2011.10.1063/1.3524336}.
See \cref{sec:methods} for details regarding the basis set and extrapolation procedures.}
\label{tab:cohesive}
\begin{tabular}{lSSSSSSS}
\toprule
{\multirow{3}{*}{System}} & \multicolumn{7}{c}{Cohesive energy (eV)} \\
\cmidrule(lr){2-8}
& \multicolumn{2}{c}{HF}
& \multicolumn{2}{c}{MP2}
& \multicolumn{2}{c}{CCSD}
& {\multirow{2}{*}{Expt.}}  \\
\cmidrule(lr){2-3}
\cmidrule(lr){4-5}
\cmidrule(lr){6-7}
& \multicolumn{1}{c}{PP} & \multicolumn{1}{c}{AE} & \multicolumn{1}{c}{PP} & \multicolumn{1}{c}{AE-FC} & \multicolumn{1}{c}{PP} & \multicolumn{1}{c}{AE-FC}  \\
\midrule
MgO	 &  3.61  &  3.63  &  5.47 &  5.52  &  5.09  &  5.10  &  5.19  \\
LiCl &  2.70  &  2.71  &  3.70 &  3.72  &  3.52  &  3.55  &  3.58  \\
LiF	 &  3.32  &  3.34  &  4.56 &  4.61  &  4.38  &  4.50  &  4.46  \\
LiH	 &  1.79  &  1.79  &  2.38 &  2.39  &  2.45  &  2.50  &  2.49  \\
BN	 &  4.61  &  4.71  &  7.02 &  7.15  &  6.42  &  6.54  &  6.76  \\
BP	 &  3.37  &  3.40  &  5.59 &  5.64  &  4.89  &  4.95  &  5.14  \\
Si	 &  3.02  &  3.02  &  5.06 &  5.07  &  4.46  & 4.47 &  4.70  \\
C	 &  5.13  &  5.27  &  7.82 &  7.99  &  7.10  &  7.26  &  7.55  \\
\midrule
MAE (eV) & 1.54 & 1.50 & 0.24 & 0.30 & 0.19 & 0.14 & $\cdots$ \\
MARE ($\%$) & 30.3 & 29.7 & 4.9 & 5.9 & 3.5 & 2.5 & $\cdots$ \\
\bottomrule
\end{tabular}
\end{table}

\subsection{Band gaps}
\label{sec:band gap}

We now turn to band gaps, which are harder to converge to the TDL, as discussed in \cref{sec:intro,sec:methods}.
Materials such as \ce{MgO}, \ce{LiCl}, \ce{LiF}, and \ce{LiH} have direct band gaps, while \ce{BN}, \ce{BP}, \ce{Si}, and \ce{C} have indirect band gaps.
We will show that TDL extrapolation is harder for indirect gaps, and for these materials, we will also compute the direct gap at the valence-band maximum (VBM; $\Gamma\rightarrow\Gamma$) for theoretical analysis.
We will first evaluate all extrapolation models using the GTH-HF pseudopotential, and then investigate the error introduced by the pseudopotential approximation at the end of this section.

\subsubsection{Direct band gaps}
\label{sec:direct}

\begin{table*}[t]
\centering
\caption{Extrapolated thermodynamic-limit (TDL) direct band gaps (in eV), obtained using different extrapolation models and fitting ranges, along with the mean absolute error (MAE) relative to the converged reference results (ABC with \meshto{3}{6}). 
For all solids, we calculate the direct band gap at the $\Gamma$ point, except for LiH, where it is calculated at the X point.
All calculations are carried out using the GTH-HF pseudopotential and the GTH-cc-pVDZ basis set.}
\label{tab:tdl_direct_gaps}
\begin{tabular}{lSSSSSSSSSSSS}
\toprule
\multirow{2}{*}{System}
 & \multicolumn{3}{c}{Band gap (eV) w/ ABC}
 & \multicolumn{6}{c}{Band gap (eV) w/ AB}
 & \multicolumn{3}{c}{Band gap (eV) w/ A} \\
\cmidrule(lr){2-4}
\cmidrule(lr){5-10}
\cmidrule(lr){11-13}
& {$2^3$--$5^3$} & {$2^3$--$6^3$} & {$3^3$--$6^3$}
& {$2^3$--$4^3$} & {$2^3$--$5^3$} & {$2^3$--$6^3$} & {$3^3$--$5^3$} & {$3^3$--$6^3$} & {$4^3$--$6^3$}
& {$3^3$--$4^3$} & {$4^3$--$5^3$} & {$5^3$--$6^3$} \\
\midrule
MgO
& 9.21 & 9.20 & 9.19
& 8.94 & 9.00 & 9.03 & 9.10 & 9.12 & 9.14
& 8.68 & 8.85 & 8.95 \\
LiCl
& 10.17 & 10.16 & 10.12
& 10.09 & 10.11 & 10.11 & 10.14 & 10.14 & 10.13
& 9.78 & 9.92 & 9.99 \\
LiF
& 15.98 & 15.95 & 15.89
& 15.89 & 15.91 & 15.92 & 15.95 & 15.94 & 15.92
& 15.60 & 15.74 & 15.80 \\
LiH
& 6.36 & 6.33 & 6.27
& 6.05 & 6.12 & 6.15 & 6.24 & 6.24 & 6.26
& 5.82 & 5.99 & 6.08 \\
BN
& 11.87 & 11.87 & 11.86
& 11.41 & 11.50 & 11.57 & 11.68 & 11.72 & 11.77
& 11.53 & 11.59 & 11.65 \\
BP
& 4.63 & 4.66 & 4.72
& 4.38 & 4.43 & 4.47 & 4.53 & 4.57 & 4.62
& 4.37 & 4.43 & 4.50 \\
Si
& 3.49 & 3.55 & 3.66
& 3.27 & 3.31 & 3.36 & 3.40 & 3.45 & 3.53
& 3.32 & 3.35 & 3.41 \\
C
& 7.69 & 7.71 & 7.74
& 7.42 & 7.48 & 7.52 & 7.58 & 7.61 & 7.66
& 7.24 & 7.37 & 7.47 \\
\midrule
MAE  & 0.07 & 0.05 & & 0.25 & 0.20 & 0.17 & 0.12 & 0.10 & 0.06 & 0.39 & 0.27 & 0.20  \\
\bottomrule
\end{tabular}
\end{table*}

\begin{table*}[t]
\centering
\caption{Extrapolated thermodynamic-limit (TDL) direct band gaps (in eV) obtained from the GW--EOM approach, along with reference EOM-CCSD values from the ABC extrapolation model with \meshto{3}{6}, as well as the corresponding differences and ratios of the fitted coefficients. For all solids, we calculate the direct band gap at the $\Gamma$ point, except for LiH, where it is calculated at the X point.
All calculations are carried out using the GTH-HF pseudopotential and the GTH-cc-pVDZ basis set.}
\label{tab:tdl_gw_eom_comparison}
\begin{tabular}{lSSSSSS}
\toprule
\multirow{2}{*}{System}
& \multicolumn{1}{c}{\multirow{2}{*}{$A/A'$}}
& \multicolumn{1}{c}{\multirow{2}{*}{$B/B'$}}
& \multicolumn{1}{c}{\multirow{2}{*}{$C/C'$}}
& \multicolumn{2}{c}{TDL band gap (eV)}
& \multicolumn{1}{c}{\multirow{2}{*}{Difference (eV)}} \\
\cmidrule(lr){5-6}
&
& 
& 
& \multicolumn{1}{c}{GW-EOM-234}
& \multicolumn{1}{c}{EOM-CCSD}
& \\
\midrule
MgO
& 1.35 & 1.40 & 1.92
& 9.31 & 9.19 & 0.13 \\
LiCl
& 1.37 & 1.48 & 35.27
& 10.24 & 10.12 & 0.12 \\
LiF
& 1.28 & 1.68 & -2.39
& 16.00 & 15.89 & 0.11 \\
LiH
& 1.59 & 1.14 & 1.73
& 6.67 & 6.27 & 0.39 \\
BN
& 1.58 & 0.85 & 1.02
& 12.27 & 11.86 & 0.41 \\
BP
& 1.76 & 0.59 & 0.73
& 5.22 & 4.72 & 0.50 \\
Si
& 1.89 & 0.57 & 0.71
& 4.07 & 3.66 & 0.42 \\
C
& 1.66 & 0.68 & 0.77
& 8.29 & 7.74 & 0.55 \\
\midrule
MAE  & & & & & & 0.33 \\
\bottomrule
\end{tabular}
\end{table*}

\begin{figure*}[t]
\centering
\includegraphics[width=6.25in]{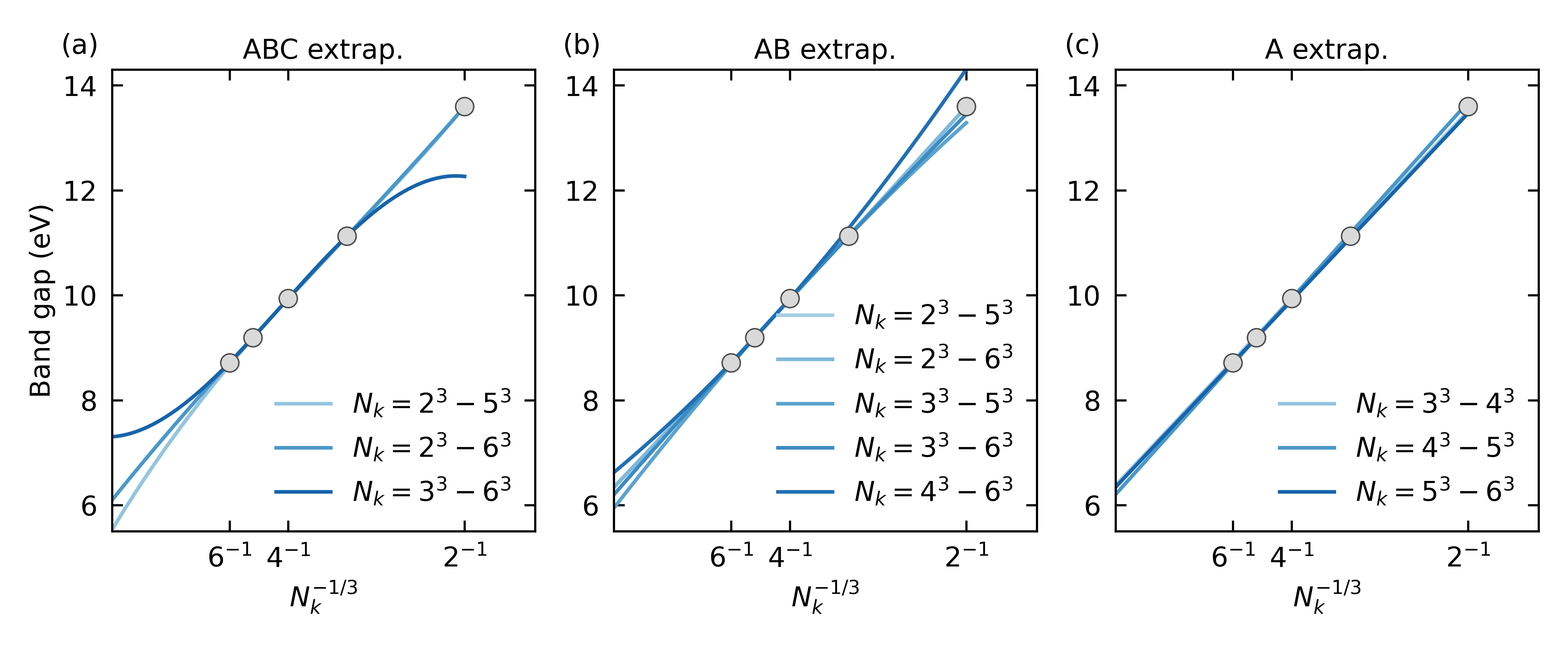} 
\caption{Extrapolation curves of the \ce{BN} indirect band gap obtained using different models and different \kmesh ranges. All calculations are carried out using the GTH-HF pseudopotential and the GTH-cc-pVDZ basis set.}
\label{fig:bn indirect band gap}
\end{figure*}

We first evaluate the different extrapolation models, performing all calculations with PPs and the double-zeta (DZ) basis.
These results are summarized in \cref{tab:tdl_direct_gaps}.
When the ABC extrapolation model is used, band gaps obtained by including \mesh{2} as the smallest \kmesh are generally close to those obtained from larger \kmesh data sets.
The largest deviation (0.17 eV) is found for the $\Gamma\rightarrow\Gamma$ gap of Si when comparing extrapolations based on the \meshto{2}{5} and \meshto{3}{6}.
Extending the former extrapolation to include the \mesh{6} point reduces this difference to 0.11 eV.

In contrast, once the $C N_k^{-1}$ term is omitted (models A and AB), extrapolations that include \mesh{2} systematically yield smaller band gaps compared to those based on larger \kmeshes. 
This behavior indicates that \mesh{2} is not sufficiently large to neglect the $C N_k^{-1}$ term in the extrapolation model, but \mesh{3} is.
For AB extrapolations starting from \mesh{3}, the largest deviation relative to the result obtained using \meshto{4}{6} is 0.13 eV.
As larger \kmeshes are used, the AB extrapolations gradually converge toward the results obtained from the ABC extrapolation with a maximum \mesh{6}.
Using the ABC extrapolation with \meshto{3}{6} as the reference, the MAE of the AB extrapolations gradually decreases from 0.25 to 0.06~eV as larger \kmeshes are included.
Based on the convergence behavior of the AB extrapolations and the relatively low variance of the ABC extrapolations, we consider the ABC extrapolation with \meshto{3}{6} as converged reference results.

As shown in \cref{tab:tdl_direct_gaps}, the single-term extrapolation (model A) exhibits the slowest convergence and large deviations from the ABC results, even when larger \kmeshes are used.
Compared to the converged reference results, the MAE of the single-term extrapolation decreases from 0.39 to 0.20 eV as the \kmeshes increase from \meshto{3}{4} to \meshto{5}{6}.
This behavior demonstrates the importance of contributions from next-to-leading-order terms in the finite-size scaling of band gaps at accessible system sizes.
The single-term extrapolation consistently underestimates the band gap, suggesting that it can be regarded as a reliable lower bound for the extrapolated band gap.

Next, we analyze the GW-EOM scheme that was proposed by the authors of Ref.~\onlinecite{moerman.2025.10.1103/PhysRevB.111.L121202}.
We analyze this approach by first performing \gwhf calculations using PySCF~\cite{sun.2018.10.1002/wcms.1340, sun.2020.10.1063/5.0006074, sun.2026.10.48550/arXiv.2603.14155} with \meshto{2}{7} and safely extrapolating to the TDL (the robustness is verified by performing extrapolations using different ranges of \kmeshes).
The proportionality constant $b$ [see \cref{eq:gw-eom-ratio}] is determined by fitting EOM-CCSD and \gwhf results with \meshto{2}{4}, which we call GW-EOM-234.
When this value of $b$ is used to extrapolate the EOM-CCSD band gaps, the accuracy of the results is erratic, as shown in \cref{tab:tdl_gw_eom_comparison}.
For three solids (MgO, LiCl, and LiF), the error is below 0.15~eV.
For the other five solids, the error is 0.39--0.55~eV.

The mixed reliability of the GW-EOM-234 scheme can be understood by numerically testing its primary assumption in \cref{eq:gw-eom-ratio}. 
From independent extrapolation fits (with model ABC), we compute ratios of expansion coefficients from \gwhf and EOM-CCSD, and we find that they are not always constant for a given solid.
For materials where $A'/A \approx B'/B$, the GW-EOM-234 approach provides a good approximation, and differences from the reference EOM-CCSD band gaps are negligible. 
Although the ratio $C'/C$ is not always consistent with the other two coefficients, its impact on the extrapolated band gap is generally small because the $N_k^{-1}$ term decays more rapidly with increasing $N_k$.
In contrast, when the ratios $A'/A$ and $B'/B$ differ significantly, the GW-EOM-234 approach overestimates the band gap, indicating that the proportionality assumption does not hold in these cases.
We emphasize that the ratios and the extracted value of $b$ depend on the choice of \kmeshes used for fitting.
We expect the reliability would be worse (better) if one had access to smaller (larger) \kmeshes than used here.

We conclude this section by discussing the basis set dependence of the direct band gaps, which---in contrast to the ground-state cohesive energies---is relatively insensitive to basis set size.
In addition to the DZ direct band gaps discussed so far, we also computed gaps with the triple-zeta (TZ) basis for \meshto{2}{5} and, whenever possible, with the quadruple-zeta (QZ) basis for \meshto{2}{4}; the corresponding TZ$-$DZ and QZ$-$DZ differences are summarized in Table~S1.
In all cases, both differences are small, with absolute values below 0.09~eV.
In this work, we estimate the basis-set incompleteness error in the direct band gap as the difference between the QZ and DZ results at the largest available \kmesh.
This correction is then added to the DZ-based TDL EOM-CCSD band gap, and these corrected results are reported in \cref{sec:final}.

\subsubsection{Indirect band gaps}
\label{sec:indirect}

Next, we investigate the convergence behavior of indirect band gaps (using PPs and the DZ basis). 
We consider the indirect band gap of \ce{BN} as a representative example.
In \ce{BN}, the VBM occurs at the $\Gamma$ point, and the conduction-band minimum (CBM) occurs at the X point.
As shown in \cref{fig:bn indirect band gap}, over the \kmeshes accessible here, model A yields the most stable estimates, and the extrapolated values obtained from the neighboring \kmesh pairs are relatively close.
However, the differences are still non-negligible: the change between extrapolations based on \meshto{4}{5} and \meshto{5}{6} is 0.14 eV, which is comparable to that observed for the direct band gap extrapolation (0.06 eV).
For the direct band gap, however, model A based on \meshto{5}{6} underestimates the converged reference EOM-CCSD band gap by 0.21 eV.
This suggests a similar magnitude of underestimation for the indirect band gap and indicates that additional data points at larger \kmeshes are required to reduce the uncertainty below 0.1 eV (we will demonstrate this point later in this section).
In contrast to the direct band gap extrapolation, higher-order extrapolation models that include sub-leading terms (models AB and ABC) become significantly more sensitive to the choice of \kmeshes used, resulting in apparent overfitting and concomitant large variations in the extrapolated band gap. 

To understand the origin of this behavior, we extrapolate the ionization potential (IP) and electron affinity (EA) separately using model ABC.
The IP is evaluated at the $\Gamma$ point, and therefore the calculation does not require any \kmesh shifting.
As shown in Fig.~S1, the IP values are straightforward to extrapolate, and the extrapolated results are stable across different \kmesh ranges. 
In contrast, the EA is evaluated at the X point, which requires that the \kmesh be shifted to ensure that the X point is included. 
The resulting EA values are considerably more difficult to extrapolate and exhibit a strong sensitivity to the choice of \kmeshes used for fitting. 
Because the indirect band gap is obtained as the sum of IP and EA, it inherits this sensitivity from the EA contribution.
We attribute the slower and less systematic convergence of EA primarily to the use of shifted \kmeshes. 

We also examine the convergence behavior of the direct band gap evaluated at the X point.
In this case, both the IP and EA are individually sensitive to the choice of \kmeshes used in the extrapolation, as shown in Fig.~S2.
However, these errors tend to cancel in the band gap, and the extrapolated direct band gap is much less sensitive to the choice of \kmeshes, similar to that obtained with non-shifted \kmeshes. 
Unfortunately, even the extrapolation of the direct band gap evaluated with shifted \kmeshes is less reliable than that obtained using non-shifted \kmeshes: the maximum variation of the extrapolated direct band gap at the X point is 0.3~eV across different \kmesh ranges, whereas the corresponding variation at the $\Gamma$ point is only 0.01~eV.

\begin{figure}[t]
\centering
\includegraphics[width=3.0in]{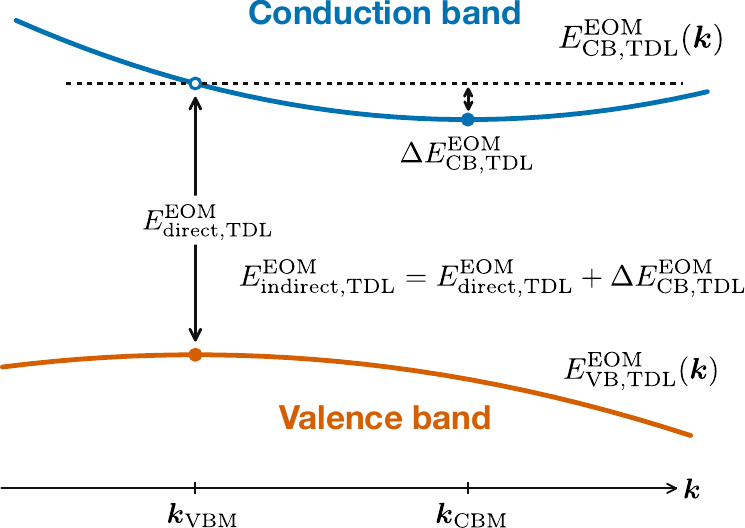} 
\caption{Illustration of the composite approach [Eqs.~(\ref{eq:composite})] used to calculate indirect band gaps.}
\label{fig:composite}
\end{figure}

Based on these findings, we propose to use a composite approach to extrapolate the EOM-CCSD indirect band gap. 
We rewrite the extrapolated indirect band gap as a sum of two contributions,
\begin{subequations}
\label{eq:composite}
\begin{gather}
\begin{split}
E_{\mathrm{indirect, TDL}}^{\mathrm{EOM}} &= E_{\mathrm{CB, TDL}}^{\mathrm{EOM}} (\bm{k}_\mathrm{CBM}) - E_{\mathrm{VB, TDL}}^{\mathrm{EOM}} (\bm{k}_\mathrm{VBM})\\
&= E_{\mathrm{direct, TDL}}^{\mathrm{EOM}} + \Delta E_{\mathrm{CB, TDL}}^{\mathrm{EOM}}
\end{split} \\
E_{\mathrm{direct, TDL}}^{\mathrm{EOM}} = E_{\mathrm{CB, TDL}}^{\mathrm{EOM}} (\bm{k}_\mathrm{VBM}) - E_{\mathrm{VB, TDL}}^{\mathrm{EOM}} (\bm{k}_\mathrm{VBM}) \\
\Delta E_{\mathrm{CB, TDL}}^{\mathrm{EOM}} = E_{\mathrm{CB, TDL}}^{\mathrm{EOM}} (\bm{k}_\mathrm{CBM}) - E_{\mathrm{CB, TDL}}^{\mathrm{EOM}} (\bm{k}_\mathrm{VBM})
\end{gather}
\end{subequations}
where, for example, $E_{\mathrm{CB, TDL}}^{\mathrm{EOM}} (\bm{k}_\mathrm{VBM})$ is the CB energy evaluated at the $k$-point where the VBM occurs, extrapolated to the TDL (this approach is illustrated graphically in Fig.~\ref{fig:composite}).
The first term, $E_{\mathrm{direct, TDL}}^{\mathrm{EOM}}$, is the direct band gap at $\bm{k}_\mathrm{VBM}$ (always the $\Gamma$ point for the indirect gap solids studied here), which is much easier to extrapolate than the indirect band gap.
The second term, $\Delta E_{\mathrm{CB, TDL}}^{\mathrm{EOM}}$ is the CB offset between $\bm{k}_\mathrm{CBM}$ and $\bm{k}_\mathrm{VBM}$.
We find that $\Delta E_{\mathrm{CB}, N_k}^{\mathrm{EOM}}$ converges more rapidly with respect to the \kmesh and can be determined with a relatively small uncertainty.

We compute the conduction-band offsets for \ce{BN}, \ce{BP}, \ce{Si}, and \ce{C} using increasingly dense \kmeshes up to the largest available size.
For C, we also evaluate the offset using an approximate value $\tilde{\bm{k}}_\mathrm{CBM}=(1/3, 0, 1/3)$, which allows for some extra savings due to space group symmetry.
For \ce{BN} and \ce{C} (with the approximate CBM), calculations are feasible up to $N_k \leq 6^3$.
For other systems, calculations are limited to $N_k \leq 5^3$ because the required \kmesh shifting reduces the available symmetry, leading to a prohibitively large number of inequivalent $k$-points for the \mesh{6} calculation. 
As shown in Table~S2, $\Delta E_{\mathrm{CB}, N_k}^{\mathrm{EOM}}$ changes by less than 0.1 eV for \kmeshes larger than \mesh{3}, and the variation is further reduced below 0.05 eV at \mesh{5}.
Therefore, we take the value of $\Delta E_{\mathrm{CB}, N_k}^{\mathrm{EOM}}$ obtained at the largest available \kmesh as a practical approximation to $\Delta E_{\mathrm{CB, TDL}}^{\mathrm{EOM}}$.

Using this composite approach, we obtain a converged reference EOM-CCSD band gap of 6.54 eV for \ce{BN} in the DZ basis.
Without this approach, extrapolation of the band gap with model A using \meshto{5}{6} predicts a band gap of 6.33 eV, underestimating the EOM-CCSD reference by 0.21 eV.
The magnitude of this underestimation is consistent with that observed for the direct band gap extrapolation, supporting our previous assumption that additional data points are required for extrapolations of indirect band gaps based on model A.
We also test the GW-EOM-234 approach, which predicts a band gap of 6.92~eV, overestimating the reference value by 0.38~eV.

We further examine the basis-set dependence of the conduction-band offsets used in the extrapolation, and the results are summarized in Table~S3.
For all systems, both the TZ$-$DZ and QZ$-$DZ differences are small, with absolute values below 0.06~eV.
In this work, the basis-set incompleteness error in the conduction-band offset is estimated as the difference between the QZ and DZ results at \mesh{3}.
This correction is then added to the DZ-based $\Delta E_{\mathrm{CB, TDL}}^{\mathrm{EOM}}$, producing the final results given in the next section.

\subsubsection{Final direct and indirect band gaps}
\label{sec:final}

Using the same TDL extrapolation and basis-set correction procedures described above, we also perform AE-FC calculations of direct and indirect band gaps.
The only exception is \ce{LiF}, for which the TZ and QZ calculations showed poor results using the settings described in \cref{sec:methods} and convergence problems when tightening the settings.
We therefore estimate the AE-FC correction to the DZ-based TDL band gap of \ce{LiF} from the difference between the AE-FC and GTH-HF DZ band gaps at \mesh{6}.
Numerical results show that this correction is nearly constant across the \kmeshes considered, ranging from 0.15 to 0.21~eV.
The basis-set incompleteness error is then estimated as the difference between the TZ and DZ results at \mesh{2}, where we were able to achieve CCSD convergence of both DZ and TZ calculations after removing diffuse primitives with exponents $< 0.05$.

We summarize our final TDL EOM-CCSD band gaps corrected for basis-set incompleteness errors in \cref{tab:band gap exp}.
We see that band gaps calculated with PPs are frequently smaller than those from AE-FC calculations by roughly 0.1~eV, a larger discrepancy than the one observed in the cohesive energies.
\cref{tab:band gap exp} also collects experimental band gaps for comparison, which have been corrected for zero-point renormalization (ZPR) collected from Refs.~\citenum{engel.2022.10.1103/PhysRevB.106.094316,moerman.2025.10.1103/PhysRevB.111.L121202}, with the ZPR values computed using many-body perturbation theory within the Allen–Heine–Cardona framework.
EOM-CCSD predicts band gaps with a MAE of 0.39~eV with PPs and a MAE of 0.42~eV without PPs.
Having eliminated finite-size and complete-basis set extrapolation errors, we believe these band gaps and MAEs to represent the converged predictions of EOM-CCSD.
The largest deviations from experiment are found for \ce{MgO}, \ce{LiH}, \ce{LiF}, and \ce{LiCl}, all of which have a rock-salt crystal structure.

To understand the origin of these relatively large deviations, we examine the single-excitation character of the IP-EOM-CCSD states, quantified by the squared norm of the amplitudes associated with the 1-hole operators in the IP calculations ($n_1^{\mathrm{IP}}$).
A previous study~\cite{moerman.2025.10.1103/PhysRevB.111.L121202} suggested that materials with $n_1^{\mathrm{IP}} > 95.5\%$ tend to exhibit small errors with respect to experiment.
We therefore analyze $n_1^{\mathrm{IP}}$ for all eight materials using the GTH-HF DZ calculations at \mesh{6}, and find that the rock-salt materials with larger errors indeed tend to have slightly smaller $n_1^{\mathrm{IP}}$ (93--94\% versus 95--95\%).
This observation highlights the limitations of EOM-CCSD for extended systems and suggests that inclusion of triple excitations may be necessary to obtain more accurate band gaps.

\begin{table}[b]
\centering
\caption{Extrapolated thermodynamic-limit (TDL) EOM-CCSD band gaps ($E_g$, in eV) obtained with GTH-HF pseudopotentials (PP) or the all-electron (AE) and frozen-core (FC) approximation, with basis-set incompleteness corrections applied. Also shown are the mean absolute error (MAE) and mean absolute relative error (MARE). Experimentally observed (obs.) and zero-point corrected values are collected from Refs.~\citenum{chiang1989electronic,madelung2004semiconductors,levinshtein2001properties,woo.2016.10.1088/2053-1591/3/7/074003,baroni.1985.10.1103/PhysRevB.32.4077,piacentini.1976.10.1103/PhysRevB.13.5530,baldini.1970.10.1002/pssb.19700380132,whited.1973.10.1016/0038-10987390754-0,engel.2022.10.1103/PhysRevB.106.094316,moerman.2025.10.1103/PhysRevB.111.L121202}. The single-excitation characters of the IP-EOM-CCSD states ($n_1^{\mathrm{IP}}$) are evaluated from the GTH-HF DZ calculations at \mesh{6}.}
\label{tab:band gap exp}
\begin{tabular}{lSSSSSSc}
\toprule
\multirow{2}{*}{System}
& \multicolumn{2}{c}{PP} & \multicolumn{2}{c}{AE}
& \multicolumn{2}{c}{Expt.\ $E_g$}
& \multirow{2}{*}{$n_1^{\mathrm{IP}}$ (\%)} \\
\cmidrule(lr){2-3}
\cmidrule(lr){4-5}
\cmidrule(lr){6-7}
& \multicolumn{1}{c}{$E_g$} 
& \multicolumn{1}{c}{Error} 
& \multicolumn{1}{c}{$E_g$} 
& \multicolumn{1}{c}{Error} 
& \multicolumn{1}{c}{corrected}
& \multicolumn{1}{c}{obs.} 
& \\
\midrule
MgO  &  9.22 &  0.86 &  9.33 &  0.97 & 8.36 & 7.83 & 93.9 \\
LiCl & 10.18 &  0.24 & 10.36 &  0.42 & 9.94 & 9.40 & 94.7 \\
LiF  & 15.93 &  0.50 & 16.03 &  0.60 & 15.43 & 14.20 & 94.6\\
LiH  &  6.28 &  0.85 &  6.21 &  0.78 & 5.43 & 4.99 & 93.6 \\
BN   &  6.62 &  0.12 &  6.73 &  0.23 & 6.50 & 6.10 & 95.4 \\
BP   &  2.05 & -0.21 &  2.07 & -0.19 & 2.26 & 2.16 & 95.7 \\
Si   &  1.29 &  0.06 &  1.23 & -0.00 & 1.23 & 1.17 & 95.1 \\
C    &  5.55 & -0.25 &  5.63 & -0.17 & 5.80 & 5.48 & 96.3\\
\midrule
MAE (eV)  &  & 0.39 &  & 0.42 &  & \\
MARE ($\%$) &  & 6.49 &  & 6.12 &  &  & \\
\bottomrule
\end{tabular}
\end{table}

\subsection{Rutile Titanium Dioxide}
As a new application of our performant implementation, we aim to calculate the band gap of rutile \ce{TiO2}.
Semiconducting metal oxides such as \ce{TiO2} are popular photocatalysts for light-driven water splitting~\cite{schneider.2014.10.1021/cr5001892, wang.2020.10.1021/acs.chemrev.9b00201}, and narrowing the band gap of \ce{TiO2} to extend its absorption edge into the visible light range represents a particular challenge in photocatalyst design. 
The experimental onset of absorption is around 3~eV~\cite{scanlon.2013.10.1038/nmat3697}, and the band gap as measured by photoemission is around 3.6~eV~\cite{rangan.2010.10.1021/jp909320f}, where the difference is partially attributable to excitonic effects.
Including the zero-point renormalization (ZPR) of $-0.31$ eV~\cite{wu.2018.10.1021/acs.jpcc.8b06941} suggests a ZPR-corrected band gap of about 3.9~eV.
DFT with the PBE functional predicts a band gap of 1.8~eV, while a larger gap of 2.9~eV is obtained at the PBE + $U$ level (with $U = 8$~eV)~\cite{arroyo-dedompablo.2011.10.1063/1.3617244}.
G$_0$W$_0$ calculations predict a gap of 3.4--3.7~eV, depending on the reference functional~\cite{kang.2010.10.1103/PhysRevB.82.085203,landmann.2012.10.1088/0953-8984/24/19/195503}.
In this section, we target the band gap of bulk rutile \ce{TiO2} with EOM-CCSD.

The structure of rutile \ce{TiO2} is obtained from the Materials Project~\cite{jain.2013.10.1063/1.4812323} with lattice parameters of 4.5998~\AA{} and 2.9592~\AA{}.
We calculate the direct band gap at the $\Gamma$ point (DFT calculations suggest that the indirect band gap $\Gamma \rightarrow \mathrm{R}$ is almost degenerate). 
We perform calculations with $2\times2\times3$ and $3\times3\times4$ \kmeshes using the GTH-HF pseudopotential and the GTH-DZVP-MOLOPT-SR basis, which give band gaps of 7.22 and 6.25~eV, respectively. 
An EOM-CCSD calculation with the $2\times2\times3$ \kmesh using the GTH-cc-pVTZ basis gives a band gap of 7.31 eV, suggesting the basis set incompleteness error is approximately 0.09 eV.
Using a two-point extrapolation based on model A and correcting for the basis-set incompleteness error predicts a band gap of 4.17 eV in the TDL, which overestimates the ZPR-corrected experimental band gap by around 0.3~eV, consistent with the accuracy we see for other solids.
As we have shown in this work, there may be some residual extrapolation error in our prediction, which typically results in an underestimation of the converged result.
However, given the larger unit cell, the anisotropic \kmeshes, and the presence of transition-metal atoms with 3d electrons, it is unclear how the trends demonstrated in the present paper translate to \ce{TiO2}.
Aside from extrapolation, the remaining error may be attributed to the low single-excitation character of the IP-EOM-CCSD state, for which we obtain  $n_1^{\mathrm{IP}} = 93.7\%$ in the $2\times2\times3$ \kmesh calculation with the DZ basis.

\section{Discussion}

Our optimized, parallel implementation of periodic CCSD and EOM-CCSD allows the largest calculations performed to date based on canonical (delocalized) orbitals, which we have used to resolve basis-set and finite-size errors.
Cohesive energies can be straightforwardly converged due to the fast decay of the finite-size error.
In Table~S4, we compare our cohesive energies to previously reported values calculated with HF, MP2, and CCSD using PySCF~\cite{ goldzak.2022.10.1063/5.0119633}, VASP~\cite{gruneis.2010.10.1063/1.3466765,booth.2013.10.1038/nature11770}, and FHI-aims~\cite{zhang.2019.10.1088/1367-2630/aaf751}.
Reassuringly, we find agreement to better than 0.05~eV for almost all solids and levels of theory, with a maximum deviation of 0.14~eV.

As we have stressed, eliminating finite-size errors in EOM-CCSD band gaps is much harder.
With access to \kmeshes up to \mesh{6}, we believe our band gaps are converged to around 0.1~eV.
With limited access to more accessible \kmeshes up to \mesh{4}, we have shown that simple two-point extrapolation uniformly underestimates the band gap by 0.3--0.5~eV, in agreement with recent work~\cite{moerman.2025.10.1103/PhysRevB.111.L121202}.
If supplemented with a large and converging set of GW band gap data, the extrapolation scheme proposed in Ref.~\citenum{moerman.2025.10.1103/PhysRevB.111.L121202} improves the prediction, but overestimates the band gap by about 0.1--0.3~eV, depending on the material.
Given that most practical calculations on three-dimensional solids are limited to \mesh{4}, we can compare three extrapolation schemes: the AB model fitted using \meshto{2}{4}, the A model fitted using \meshto{3}{4}, and the GW-EOM-234 model. 
As shown in \cref{tab:tdl_gw_eom_comparison,tab:tdl_direct_gaps}, the AB model is the most accurate among these approaches, giving a MAE of 0.25 eV relative to the converged results, and it is therefore the extrapolation approach we recommend in such cases.
In Table~S5, we compare our EOM-CCSD band gaps to previously reported values using the same three codes; unlike for cohesive energies, we observe discrepancies around 0.3~eV on average, consistent with our analysis of existing extrapolation schemes.

Given the robustness of our own extrapolations, which we estimate are converged to about 0.1~eV, we arrive at an average band gap error of EOM-CCSD compared to experiment of around 0.3--0.4~eV (typically an overestimate), with errors as large as 0.8--1~eV for ionic solids like LiH or MgO.
These large errors are worthy of further study and may be attributable to an insufficient treatment of electron correlation, electron-phonon interactions, or both. 

\section{Methods}
\label{sec:methods}

\subsection{Implementation}

Our new implementation of ground-state CCSD and EOM-CCSD is based on PySCF.
It combines distributed and shared memory parallelism and exploits space-group symmetry, which reduces computational effort by working with the irreducible Brillouin zone.
We summarize the background theory for periodic CCSD and EOM-CCSD in Appendix~\ref{app:ccsd}, and we describe our parallel implementation in Appendix~\ref{app:mpi}.
Executing our code on up to 12 nodes, each with 96 cores and 1.5~TB of memory per node, we are able to sample the Brillouin zone with up to $N_k = 6^3$ in a double-zeta (DZ) basis set, up to $N_k = 5^3$ in a triple-zeta (TZ) basis set, and up to $N_k = 4^3$ in a quadruple-zeta (QZ) basis set without any approximations, as described more below.

\subsection{Computational details}

We study eight semiconductors and insulators with different crystal structures, including diamond (\ce{C}, \ce{Si}), zinc blende (\ce{BN}, \ce{BP}), and rock salt (\ce{MgO}, \ce{LiH}, \ce{LiF}, \ce{LiCl}).
All calculations are performed with experimental lattice constants listed in \Cref{tab:parameter}.
The Brillouin zone is sampled with $N_k$ $k$-points from a uniform mesh;
except where explicitly stated, all \kmeshes include the $\Gamma$ point.
Gaussian density fitting is used to evaluate and compress the two-electron repulsion integrals of the Ewald interaction~\cite{sun.2017.10.1063/1.4998644, ye.2021.10.1063/5.0046617}.
We use Madelung constant corrections (also known as the probe-charge Ewald method~\cite{paier.2005.10.1063/1.1926272, broqvist.2009.10.1103/PhysRevB.80.085114, sundararaman.2013.10.1103/PhysRevB.87.165122} for HF exchange) to eliminate leading-order finite-size errors wherever possible.
With this correction, both the HF ground-state energy and the HF orbital energies converge to the TDL asymptotically as $N_k^{-1}$.
We also calculate ground-state energies using second-order M{\o}ller-Plesset perturbation theory (MP2) and CCSD, whose correlation energies show the same $N_k^{-1}$ convergence behavior.

We perform all-electron (AE) HF calculations and apply the frozen core (FC) approximation for subsequent MP2 and CCSD calculations. 
We use the correlation-consistent cc-pV$X$Z ($X = \mathrm{D}, \mathrm{T}, \mathrm{Q}$) basis set~\cite{dunning.1989.10.1063/1.456153a}.
When used in compact solids, these basis sets, which were optimized for isolated atoms and contain diffuse functions, can exhibit large linear dependencies, which we resolve via canonical orthogonalization, discarding overlap eigenvalues less than $10^{-6}$.
For solids containing Li and Mg, we additionally discard primitive Gaussians with exponents smaller than 0.05.
For \ce{LiF} and \ce{LiH} with the TZ and QZ basis sets, a more aggressive cutoff (discarding exponents $<$ 0.10) is used in a separate set of HF and CCSD calculations, from which the correlation energy is added to the HF energy obtained without this more aggressive cutoff.
Numerical tests on small \kmeshes show that the error introduced by the frozen-core approximation is consistently below 0.05~eV for both cohesive energies and band gaps, although we did not test the use of larger core-valence basis sets.
We also perform calculations with GTH pseudopotentials (PPs) optimized for Hartree-Fock theory (GTH-HF-rev)~\cite{goedecker.1996.10.1103/PhysRevB.54.1703a,hartwigsen.1998.10.1103/PhysRevB.58.3641} together with the correlation-consistent GTH-cc-pV$X$Z ($X = \mathrm{D}, \mathrm{T}, \mathrm{Q}$) basis sets~\cite{ye.2022.10.1021/acs.jctc.1c01245}; importantly, these basis sets were optimized for use in compact solids like the ones studied here, which facilitates basis set convergence.

\subsection{Cohesive energy calculations and extrapolation}

\begin{figure*}[t]
\centering
\includegraphics[width=6.25in]{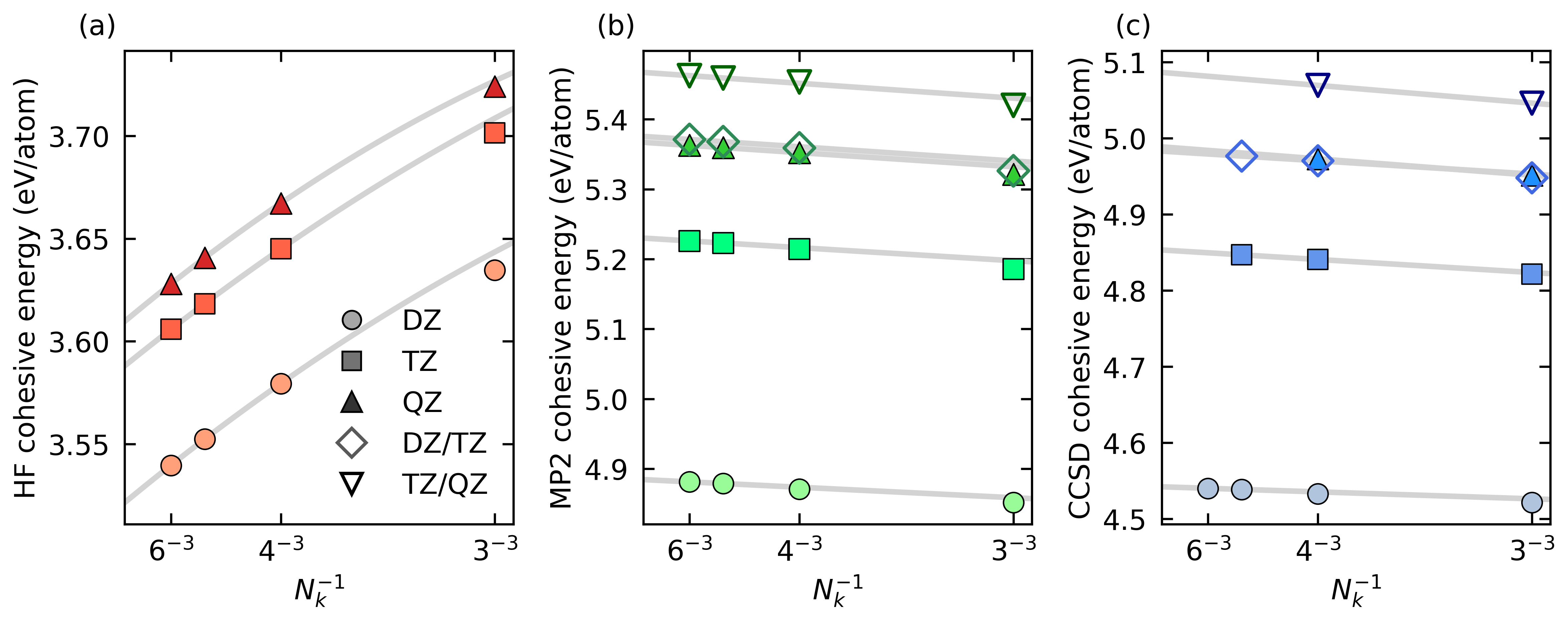} 
\caption{Thermodynamic-limit convergence of the cohesive energies of \ce{MgO} using the GTH pseudopotentials and the GTH-cc-pV$X$Z ($X = \mathrm{D}, \mathrm{T}, \mathrm{Q}$) basis sets. For each \kmesh, the MP2 and CCSD correlation energies are added to a separately converged HF energy. We also report results obtained using MP2 and CCSD correlation energies extrapolated to the complete-basis-set limit (from TZ/QZ using an $X^{-3}$ form).}
\label{fig:mgo cohesive}
\end{figure*}

For cohesive energies, the atomic energies of open-shell atoms are calculated using unrestricted HF, MP2, and CCSD, and basis-set superposition error is accounted for by including ghost atoms with crystalline basis functions.
For atomic CCSD calculations in the TZ and QZ basis, we employ the frozen natural orbital approximation~\cite{sosa.1989.10.1016/0009-26148987399-3a,taube.2005.10.1135/cccc20050837,landau.2010.10.1063/1.3276630,deprince.2013.10.1021/ct300780u,deprince.2013.10.1021/ct400250ua} with a conservative NO occupation number cutoff of $10^{-6}$ to reduce the computational cost.

As an example of a ground-state workflow, in \cref{fig:mgo cohesive}, we show the predicted cohesive energy of \ce{MgO} using $k$-point meshes up to $N_k = 6^3$ and the GTH PPs and basis sets.
Following previous work~\cite{goldzak.2022.10.1063/5.0119633}, the HF energies are extrapolated according to a three-parameter fit with sub-leading corrections, while the MP2 and CCSD correlation energies are extrapolated using a simpler two-parameter fit with only leading-order finite-size errors,
\begin{subequations}
\begin{align}
E_{\mathrm{HF}} (N_k) &= E_{\mathrm{HF}} (\infty) + A N_k^{-1} 
    + B N_k^{-2}, \\
E_\mathrm{c} (N_k) &= E_\mathrm{c} (\infty) + A N_k^{-1}.
\end{align}
\end{subequations}

\begin{table}[b]
\centering
\caption{Lattice parameters ($a$) and positions of valence-band maximum (VBM) and conduction-band minimum (CBM) in reciprocal space in relative coordinates for the materials studied in this work.}
\label{tab:parameter}
\begin{tabular}{lccc}
\toprule
System
& $a$ (\AA{})
& $\bm{k}_{\mathrm{VBM}}$
& $\bm{k}_{\mathrm{CBM}}$\\
\midrule
MgO  & 4.213 & (0.0, 0.0, 0.0) & (0.0, 0.0, 0.0) \\
LiCl & 5.130 & (0.0, 0.0, 0.0) & (0.0, 0.0, 0.0) \\
LiF  & 4.035 & (0.0, 0.0, 0.0) & (0.0, 0.0, 0.0) \\
LiH  & 4.083 & (0.5, 0.0, 0.5) & (0.5, 0.0, 0.5) \\
BN   & 3.615 & (0.0, 0.0, 0.0) & (0.5, 0.0, 0.5) \\
BP   & 4.538 & (0.0, 0.0, 0.0) & (0.4114, 0.0, 0.4114)\\
Si   & 5.431 & (0.0, 0.0, 0.0) & (0.4193, 0.0, 0.4193) \\
C    & 3.567 & (0.0, 0.0, 0.0) & (0.3646, 0.0, 0.3646) \\
\bottomrule
\end{tabular}
\end{table}

For HF, we fit the data over the range \meshto{4}{6} and for MP2 and CCSD, we fit over the range \meshto{5}{6};
for CCSD with the TZ basis, the fitting range is reduced to \meshto{4}{5}, and for CCSD with the QZ basis to \meshto{3}{4}, both due to computational limitations.
We estimate the reliability of these extrapolations by fitting different ranges of data points, which confirms that the extrapolations are accurate to approximately 0.01 eV.
For a simple insulator such as \ce{MgO}, we see fast convergence of the cohesive energy to the TDL. 
Our new implementation (which allows calculations up to $N_k = 6^3$) enables us to confirm that a two-parameter fit of the correlation energy is a reliable approximation, even when fitting over the relatively small range \meshto{3}{4}.

We next examine convergence of the cohesive energy to the complete basis set (CBS) limit.
As seen in \cref{fig:mgo cohesive}, the HF cohesive energies calculated with the TZ and QZ basis sets differ by only 0.02~eV.
We consider the QZ values to be sufficiently close to the CBS limit, and we will use these throughout this work.  
For the MP2 and CCSD correlation energies, which show a stronger basis-set dependence and are more expensive to evaluate in a large basis set, we compare the $X^{-3}$ extrapolation results obtained from DZ/TZ and TZ/QZ pairs.
These extrapolated values differ from one another by less than 0.1~eV for MP2 and CCSD at all available $k$-point meshes.
We henceforth use TZ/QZ extrapolation for MP2 and CCSD cohesive energies.

For AE-FC calculations, we use the same TDL and CBS extrapolation procedures described above.
Our numerical tests indicate that the resulting errors are below 0.05~eV for all systems except \ce{MgO}, for which the DZ/TZ- and TZ/QZ-based CBS extrapolations differ by about 0.5~eV.
An additional quintuple-zeta (5Z) calculation with \mesh{2} shows that the DZ/TZ extrapolation overestimates the MP2 and CCSD 5Z cohesive energies by only about 0.05~eV, whereas the TZ/QZ extrapolation overestimates them by about 0.4~eV.
Therefore, for AE-FC calculations of \ce{MgO}, we adopt DZ/TZ extrapolation.

\subsection{Band gap calculations and extrapolation}

\begin{figure*}[t]
\centering
\includegraphics[width=6.25in]{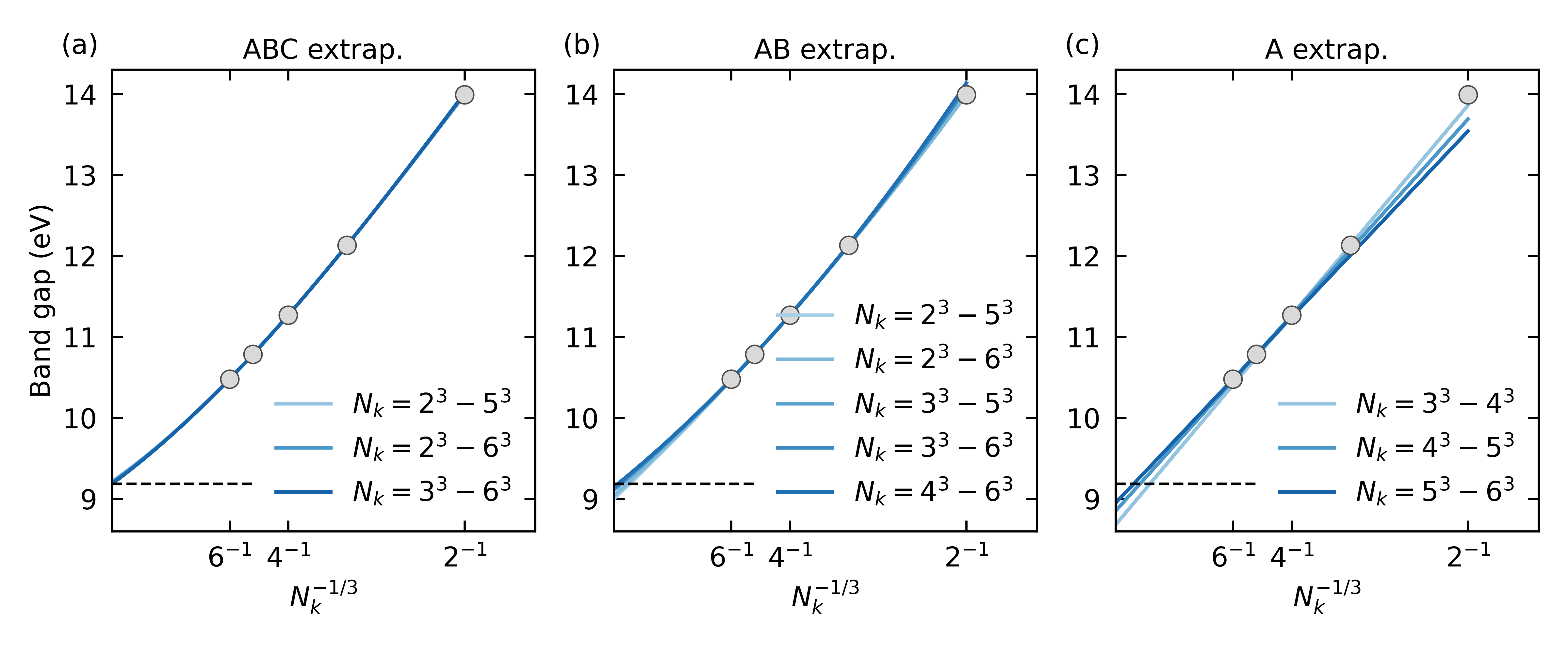} 
\caption{Extrapolation curves of the \ce{MgO} direct band gap obtained using different models and different \kmesh ranges. All calculations are carried out using the GTH pseudopotentials and the GTH-cc-pVDZ basis set. The black dashed line represents the value obtained from the ABC extrapolation model using \meshto{3}{6}. }
\label{fig:mgo direct}
\end{figure*}

For band gaps, we compute the ionization potential (IP) and electron affinity (EA) using EOM-CCSD, defined as $E(N-1) - E(N)$ and $E(N+1) - E(N)$, respectively, where $E(N)$ is the ground-state energy of an $N$-electron system.
The band gap is then obtained as the sum of the IP and EA.
When necessary, IP and EA calculations are performed using shifted \kmeshes to include the $k$-points corresponding to the valence band maximum (VBM) or conduction band minimum (CBM).
The $k$-points corresponding to the VBM and CBM are determined from DFT calculations using the PBE functional with a \mesh{6} \kmesh and are listed in \cref{tab:parameter}.

As discussed in \cref{sec:intro}, band gaps converge to the TDL slower than ground-state energies, precluding simple extrapolation.
Given the large \kmeshes accessible with our new implementation, we are able to evaluate fits with up to four parameters~\cite{moerman.2025.10.1021/acs.jctc.4c01451},
\begin{equation}
\label{eq:eom extrapolation}
E_{\mathrm{gap}}^{\mathrm{CC}} (N_k) = E_{\mathrm{gap}}^{\mathrm{CC}} (\infty) + A N_k^{-1/3} +  B N_k^{-2/3} + C N_k^{-1}.
\end{equation}
In this work, we consider three models: model ABC, which retains terms up to third order; model AB, which retains terms up to second order; and model A, which retains only the leading-order term and is equivalent to the extrapolation scheme used in many previous works~\cite{mcclain.2017.10.1021/acs.jctc.7b00049,gao.2020.10.1103/PhysRevB.101.165138,vo.2024.10.1063/5.0187856} with relatively small \kmeshes.
By varying the range over which these fits are performed, we evaluate the robustness of the TDL prediction and provide our best estimates as new reference values. 
Again, an example workflow is shown for MgO (with PPs and the DZ basis set) in \cref{fig:mgo direct}.
We see that when all three fitting models are applied to the largest \kmeshes accessible by our implementation, the extrapolated band gaps agree to within 0.25~eV.

In recent work, the authors of Ref.~\citenum{moerman.2025.10.1103/PhysRevB.111.L121202} suggested that band gaps from the more affordable GW approximation converge with the same functional form,
\begin{equation}
\label{eq:gw extrapolation}
E_{\mathrm{gap}}^{\mathrm{GW}} (N_k) = E_{\mathrm{gap}}^{\mathrm{GW}} (\infty) + A' N_k^{-1/3} +  B' N_k^{-2/3} + C' N_k^{-1}.
\end{equation}
If the expansion coefficients are proportional to one another, with the same proportionality constant,
\begin{equation}
\label{eq:gw-eom-ratio}
A/A' \approx B/B' \approx C/C' = b,
\end{equation}
then the finite-size behavior of the EOM-CCSD band gap can be rewritten as
\begin{equation}
\label{eq: gw-eom}
    E_{\mathrm{gap}}^{\mathrm{CC}} (N_k) = E_{\mathrm{gap}}^{\mathrm{CC}} (\infty) + b \times \left[E_{\mathrm{gap}}^{\mathrm{GW}} (N_k) - E_{\mathrm{gap}}^{\mathrm{GW}} (\infty)\right],
\end{equation}
and the proportionality constant $b$ can be determined by fitting EOM-CCSD and GW results using a range of relatively small \kmeshes.
Using this approach, these authors predicted band gaps in the TDL using EOM-CCSD calculations only up to $N_k=3^3$ (using VASP~\cite{kresse.1994.10.1088/0953-8984/6/40/015, kresse.1996.10.1103/PhysRevB.54.11169}) or up to $N_k=4^3$ (using FHI-aims~\cite{blum.2009.10.1016/j.cpc.2009.06.022, moerman.2022.10.21105/joss.04040}) for the same eight solids studied here.
We will evaluate this approach using our own \gwhf calculations performed with PySCF~\cite{zhu.2021.10.1021/acs.jctc.0c00704} using \meshto{2}{7}. 
Following Ref.~\onlinecite{moerman.2025.10.1103/PhysRevB.111.L121202}, all \gwhf calculations are carried out without head-and-wing corrections~\cite{hybertsen.1987.10.1103/PhysRevB.35.5585}.

\section*{Data Availability}
All data supporting the findings of this work are available at \url{https://github.com/shuhangli98/supporting_data}.

\section*{Code Availability}
The software used to generate the data presented in this work is available at \url{https://github.com/shuhangli98/mpi_periodic_eom_ccsd}.

\begin{acknowledgments}
This work was supported in part by the U.S. Department of Energy under Award DE-SC0024532 (S.L. and F.A.E.) and the U.S. National Science Foundation under Award OAC-2513476 (T.C.B.).
The work of S.L. was supported in part by a Flatiron Institute predoctoral fellowship.
The Flatiron Institute is a division of the Simons Foundation.
\end{acknowledgments}

\section*{Author Contributions}
S.L. developed the software, performed the calculations, analyzed the data, prepared the figures, and wrote the original draft. 
T.C.B. conceived and supervised the project and contributed to the data analysis and manuscript preparation.
H.Z. provided conceptual input on the software and contributed to scientific discussions.
F.A.E. contributed to scientific discussions.
H.Z. and F.A.E. reviewed and edited the manuscript.
All authors reviewed and approved the final manuscript.

\section*{Competing Interests}
The authors declare no competing interests.

\appendix
\section{Periodic CCSD and EOM-CCSD}
\label{app:ccsd}
The coupled-cluster ansatz parameterizes the ground-state wavefunction with an exponential operator, $e^{\hat{T}}$, acting on the single-determinant reference state $\ket{\Phi_0}$,
\begin{equation}
    \ket{\Psi_0} = e^{\hat{T}}\ket{\Phi_0}.
\end{equation}
For systems with translational symmetry, the hole and particle spin orbitals carry crystal-momentum labels, and consequently the CCSD cluster operators can be written as
\begin{equation}
    \begin{split}
        \hat{T}_1 &= \sum'_{\boldsymbol{k}_i\boldsymbol{k}_a}\sum_{ia}t_{i\boldsymbol{k}_i}^{a\boldsymbol{k}_a}\hat{a}_{a\boldsymbol{k}_a}^{\dagger}\hat{a}_{i\boldsymbol{k}_i}^{\vphantom{\dagger}}, \\
        \hat{T}_2 &= \frac{1}{4}\sum'_{\boldsymbol{k}_i\boldsymbol{k}_j\boldsymbol{k}_a\boldsymbol{k}_b}\sum_{ijab}t_{i\boldsymbol{k}_i j\boldsymbol{k}_j}^{a\boldsymbol{k}_a b\boldsymbol{k}_b}\hat{a}_{a\boldsymbol{k}_a}^{\dagger}\hat{a}_{b\boldsymbol{k}_b}^{\dagger}\hat{a}_{j\boldsymbol{k}_j}^{\vphantom{\dagger}}\hat{a}_{i\boldsymbol{k}_i}^{\vphantom{\dagger}},
    \end{split}
\end{equation}
where $i,j$ label hole (occupied) spin orbitals, $a,b$ label particle (virtual) spin orbitals, and $\boldsymbol{k}_i, \boldsymbol{k}_j, \boldsymbol{k}_a, \boldsymbol{k}_b$ denote the corresponding crystal momenta. 
The primed summations indicate conservation of crystal momentum, $\boldsymbol{k}_i+\boldsymbol{k}_j-\boldsymbol{k}_a-\boldsymbol{k}_b = \boldsymbol{G}$, where $\boldsymbol{G}$ is a reciprocal lattice vector.

Excited-state wavefunctions can be obtained through the equation-of-motion (EOM) formalism.
In EOM-CC, excited states are described by the ansatz $\ket{\Psi_{\alpha}} = \hat{R}_{\alpha}\ket{\Psi_0}$, where $\hat{R}_{\alpha}$ is an excitation operator that generates the $\alpha$th excited state from the CCSD ground-state wavefunction. 
Henceforth, we will omit the $\alpha$ index for simplicity.
The corresponding excitation energies and states are obtained by diagonalizing the similarity-transformed Hamiltonian in the chosen excitation space.
In this work, we compute the ionization potential (IP) and electron affinity (EA) at the EOM-CCSD level of approximation.
The IPs are computed in the space of 1-hole (1h) and 2-hole, 1-particle (2h1p) states, while the EAs are computed in the space of 1-particle (1p) and 2-particle, 1-hole (2p1h) states.
Specifically, for periodic systems,
\begin{align}
    \hat{R}^{-}(\boldsymbol{k}) &= \sum_{i}r_{i\boldsymbol{k}}\hat{a}_{i\boldsymbol{k}} + \frac{1}{2}\sum'_{\boldsymbol{k}_i\boldsymbol{k}_j\boldsymbol{k}_b}\sum_{ijb}r_{i\boldsymbol{k}_ij\boldsymbol{k}_j}^{\phantom{a\boldsymbol{k}_a}b\boldsymbol{k}_b}\hat{a}_{b\boldsymbol{k}_b}^{\dagger}\hat{a}_{j\boldsymbol{k}_j}^{\vphantom{\dagger}}\hat{a}_{i\boldsymbol{k}_i}^{\vphantom{\dagger}}, \\
    \hat{R}^{+}(\boldsymbol{k}) &= \sum_{a}r_{a\boldsymbol{k}}\hat{a}_{a\boldsymbol{k}}^{\dagger} + \frac{1}{2}\sum'_{\boldsymbol{k}_j\boldsymbol{k}_a\boldsymbol{k}_b}\sum_{jab}r_{\phantom{i\boldsymbol{k}_i}j\boldsymbol{k}_j}^{a\boldsymbol{k}_a b\boldsymbol{k}_b}\hat{a}_{a\boldsymbol{k}_a}^{\dagger}\hat{a}_{b\boldsymbol{k}_b}^{\dagger}\hat{a}_{j\boldsymbol{k}_j}^{\vphantom{\dagger}},
\end{align}
where $\hat{R}^{-}(\boldsymbol{k})$ and $\hat{R}^{+}(\boldsymbol{k})$ are excitation operators for IP and EA, respectively, which create ionized states with a total momentum $\boldsymbol{k}$.

\section{Parallel periodic CCSD and EOM-CCSD algorithm}
\label{app:mpi}

In this section, we describe the memory-distributed parallel implementation of periodic CCSD and EOM-CCSD.
The dominant cost in periodic EOM-CCSD comes from the ground-state CCSD computation. 
The computational cost of periodic CCSD scales as $\mathcal{O}(N_k^4N_\mathrm{occ}^2N_\mathrm{vir}^4)$, where $N_k$ is the number of $k$-points sampled in the Brillouin zone, and $N_\mathrm{occ}$ and $N_\mathrm{vir}$ denote the number of occupied and virtual orbitals per unit cell, respectively.
For EOM-CCSD calculations at a given $k$-point, the computational scaling of the IP and EA calculations is $\mathcal{O}(N_k^3N_\mathrm{occ}^3N_\mathrm{vir}^2)$ and $\mathcal{O}(N_k^3N_\mathrm{occ}N_\mathrm{vir}^4)$, respectively.
In terms of memory usage, two-electron quantities such as the CCSD $t_2$-amplitudes are the largest objects.
Large blocks are either computed on-the-fly or distributed across MPI processes, as described below.

\subsection{Space-group symmetry}

For a dense mesh with $N_k$ $k$-points, two-electron quantities such as $t_{ij}^{ab}$ and the two-electron integrals are indexed by four $k$-points $(\boldsymbol{k}_i, \boldsymbol{k}_j, \boldsymbol{k}_a, \boldsymbol{k}_b)$, of which only three are independent due to crystal momentum conservation $\boldsymbol{k}_i+\boldsymbol{k}_j-\boldsymbol{k}_a-\boldsymbol{k}_b = \boldsymbol{G}$, where $\boldsymbol{G}$ is a reciprocal lattice vector.
Exploiting the space-group symmetry of the crystal reduces the number of independent $k$-quartets from $N_k^3$ to $N_k^3 / \mathcal{G}$, where $\mathcal{G}$ is the order of the point group. 
This reduced set is referred to as the irreducible Brillouin zone (IBZ) quartets, which we distribute over $P$ MPI processes. 
All two-electron quantities are stored only over IBZ quartets, and any BZ quartet outside the IBZ is recovered on-the-fly by applying the appropriate symmetry rotation to the corresponding IBZ block.
For example, for an IBZ block $t_{ij}^{ab}(\mathrm{IBZ})$ with quartet $(\boldsymbol{k}_i, \boldsymbol{k}_j, \boldsymbol{k}_a, \boldsymbol{k}_b)$ and a BZ quartet related to it by a point-group operation $\hat{R}$, the BZ block is obtained via
\begin{equation}
t_{i'j'}^{a'b'}(\mathrm{BZ}) = \sum_{ijab} t_{ij}^{ab}(\mathrm{IBZ}) U_{ii'}^{\hat{R}}(\boldsymbol{k}_i) U_{aa'}^{\hat{R}*}(\boldsymbol{k}_a) U_{jj'}^{\hat{R}}(\boldsymbol{k}_j)  U_{bb'}^{\hat{R}*}(\boldsymbol{k}_b) ,
\end{equation}
where $U^{\hat{R}}(\boldsymbol{k})$ are the MO rotation matrices corresponding to $\hat{R}$ at the corresponding $k$-points. 
This transformation is implemented as a sequence of four tensor contractions with a combined cost of $\mathcal{O}(N_\mathrm{occ}^2 N_\mathrm{vir}^3)$.
The two sources of speedup, MPI parallelism and updating of IBZ quartets only, yield an effective speedup of up to $P \times \mathcal{G}$ relative to a single-process calculation based on the full BZ.
Speedup by an additional factor of $\mathcal{G}$ is in principle possible, by enforcing point group symmetry within contractions, but this is not currently implemented because the point groups are generally non-Abelian.

An analogous symmetry relationship applies for the EOM-CCSD $r_{ij}^{b}$ and $r_{j}^{ab}$ amplitudes over IBZ triplets.
However, for the systems studied here, the full BZ amplitudes are small enough to store without distribution, and we update all BZ triplets.
In principle, further savings are possible, but the timings are dominated by the ground-state CCSD calculation, as shown in Sec.~\ref{sec:performance}.

\subsection{Task scheduling and load balancing}
A naive distribution of IBZ quartets across processes leads to a severe load imbalance.
The reason is that some BZ quartets coincide with stored IBZ blocks and can therefore be accessed with negligible overhead, while others must be reconstructed from their IBZ counterparts via the explicit $\mathcal{O}(N^5)$ rotation described above.
As a result, different MPI processes will have a different number of such rotations leading to imbalanced floating-point costs.

To address the load-imbalance issue, IBZ quartet computations (tasks) are assigned to each MPI process in advance through a dry-run procedure in which the computational cost of each task is estimated without performing any actual tensor contractions.
The tensor contractions within each task have the same cost and therefore do not contribute to load imbalance.
The only source of imbalance is the number of tensor-block accesses that fall outside the IBZ, which require an $\mathcal{O}(N^5)$ rotation.
We therefore define a cost function that counts the number of indirect (out-of-IBZ) accesses encountered in the inner loops of each task, and use this count as the task weight in the task-assignment procedure.

The dry-run weights are then used to assign tasks to processes via a constrained greedy scheduling algorithm. 
Two objectives are enforced simultaneously: 1) the number of tasks per process is kept as equal as possible (differing by at most one), and 2) subject to that constraint, the total weight (estimated FLOP count) per process is balanced. 
Tasks are sorted by decreasing weight and assigned using a greedy longest-processing-time-first (LPT) strategy.
A min-heap over MPI processes tracks the current cumulative load, while a per-processes quota enforces an approximately equal number of tasks.
At each step, the least-loaded process is assigned the next task and reinserted into the heap only if its quota is not exhausted.
This balances both task count and estimated computational cost.

\subsection{Memory management}
Storing all two-electron quantities over the IBZ becomes prohibitively expensive for dense $k$-point grids.
To reduce memory usage, we store only the three-index density-fitting (DF) integrals $L^{Q}_{pq}$ (replicated across processes) and compute all two-electron integral blocks involving at least two virtual orbitals (OVOV, VOOV, VOVV, and VVVV)  on-the-fly.
Here, O and V denote the occupied and virtual orbital spaces, and the block labels follow the standard physicists' notation for two-electron integrals.
Small two-electron integrals blocks such as OOOO and OOOV, as well as all one-electron quantities, are precomputed and replicated in memory.

All CCSD and IP/EA-EOM-CCSD intermediates~\cite{hirata.2004.10.1063/1.1637577, nooijen.1995.10.1063/1.468592} are evaluated in parallel.
Intermediates with four virtual indices are computed on-the-fly, while those with two virtual indices are distributed across MPI processes so that each process stores only the blocks it owns.
The CCSD $t_2$-amplitudes are partitioned in the same manner, while all remaining amplitudes are replicated.
For each computational step, tensor blocks required by local tasks are communicated using \texttt{MPI\_Alltoallv} and temporarily stored in memory buffers.
After completing the local contractions, the updated blocks are returned to their owner processes via a second \texttt{MPI\_Alltoallv}, and the temporary buffers are released before proceeding to the next step.
To further reduce the size of memory buffers, we split the tasks assigned to each process into $K$ phases. 
Within each phase, we communicate only the tensor blocks required for that phase, and free the buffers before proceeding to the next phase.
This phase-based strategy reduces the peak memory footprint per process by approximately a factor of $K$ compared to fetching all required blocks simultaneously, at the cost of performing $K$ communication rounds per step.
The communication pattern for each phase, specifically, which IBZ blocks are required from which processes, is determined during the dry-run scheduling stage.
The peak memory usage can be reduced by increasing the number of phases, providing a controllable trade-off between memory consumption and communication frequency.

\subsection{Performance}
\label{sec:performance}

\begin{figure}[t]
\centering
\includegraphics[width=3.125in]{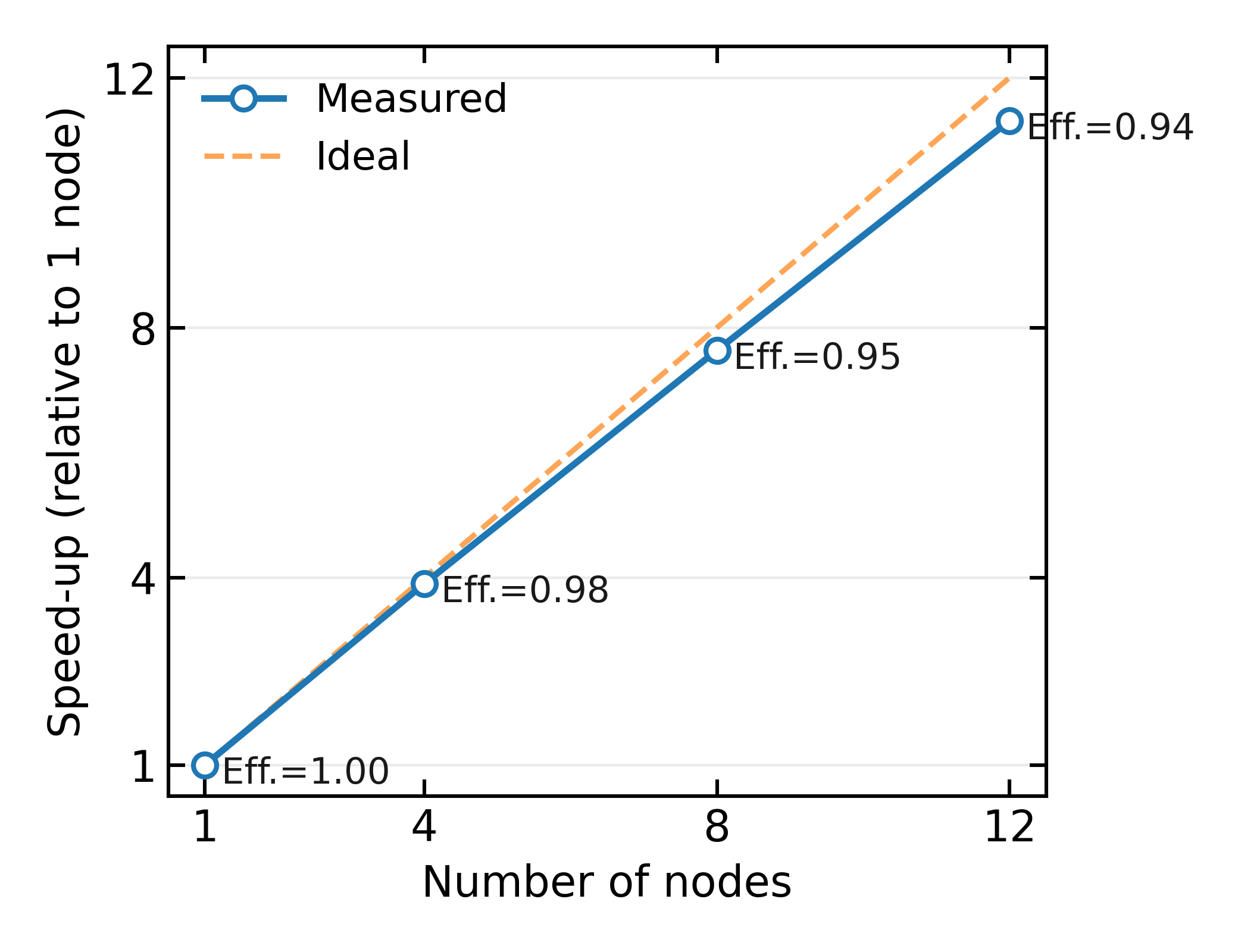} 
\caption{Speed-up of average time per CCSD iteration relative to the single-node computation. Parallel efficiencies (speed-up/nodes) are labeled next to each data point.}
\label{fig:scaling}
\end{figure}

We assess our parallel implementation using the ground-state CCSD calculation of rutile \ce{TiO2} on a $2\times2\times3$ \kmesh, employing the GTH-DZVP-MOLOPT-SR basis.
All calculations were performed on 96-core AMD Genoa nodes with 1.5 TB of memory per node.
We compute the average wall time per CCSD iteration with number of nodes from 1 to 12, and the speed-up relative to the single-node computation is shown in \cref{fig:scaling}.

The default serial PySCF implementation, which does not exploit space-group symmetry, requires about 41442 seconds per CCSD iteration.
In contrast, the PySCF serial symmetry-adapted implementation reduces this cost to about 27673 seconds per iteration.
Our implementation accelerates the serial symmetry-adapted code by optimizing the tensor-contraction routes, and it already shows a clear benefit on a single node, reducing the cost to about 17511 seconds per CCSD iteration.
Compared to the serial implementation without space-group symmetry, the reduction arises from working in the IBZ, which reduces the number of unique $k$-point quartets from 1728 (full BZ) to 504.

As shown in \cref{fig:scaling}, the wall time per CCSD iteration scales nearly linearly with the number of nodes up to 12 nodes, with speed-ups close to the ideal line.
At 12 nodes, we obtain a speed-up of 11.31.
The corresponding parallel efficiency remains high ($\geq 0.94$), indicating that communication and load imbalance introduce only minor overhead at this scale.
The calculation can be further accelerated by increasing task-level parallelism within each node.
For 12 nodes, the average CCSD iteration time is 1557 seconds with a $12\times1\times96$ configuration (nodes $\times$ tasks per node $\times$ CPUs per task) and decreases to 437 seconds with $12\times4\times24$.
Therefore, all calculations in this work use 4 tasks per node and 24 CPUs per task, with the number of nodes chosen according to system size.
Finally, we note that the CCSD, EA-EOM-CCSD, and IP-EOM-CCSD part of these calculations took $73\%$, $23\%$, and $4\%$ of the total time, respectively.

\clearpage
\bibliography{ref}

\end{document}


\title{Supplementary Information for ``Reaching the thermodynamic limit of periodic CCSD cohesive energies and band gaps''}

\author{Shuhang Li}
 \affiliation{Department of Chemistry and Cherry Emerson Center for Scientific Computation, Emory University, Atlanta, Georgia 30322, United States}
\author{Huanchen Zhai}
 \affiliation{Initiative for Computational Catalysis, Flatiron Institute, New York, New York 10010, United States}
\author{Francesco A. Evangelista}
 \affiliation{Department of Chemistry and Cherry Emerson Center for Scientific Computation, Emory University, Atlanta, Georgia 30322, United States}
\author{Timothy C. Berkelbach}
 \email{tberkelbach@flatironinstitute.org}
 \affiliation{Initiative for Computational Catalysis, Flatiron Institute, New York, New York 10010, United States}
 \affiliation{Department of Chemistry, Columbia University, New York, New York, 10027, United States}

\date{\today}

\maketitle

\begin{figure*}[!htb]
\centering
\includegraphics[width=6.25in]{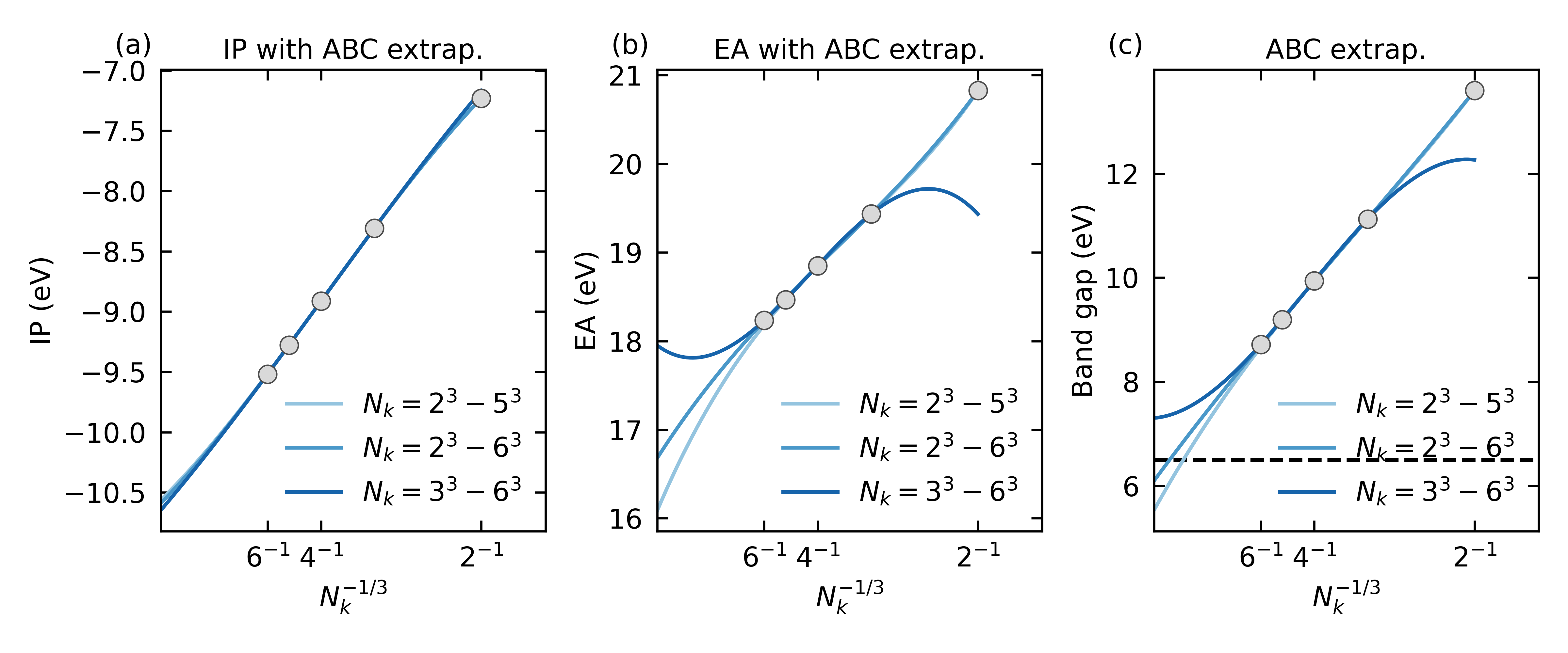}
\caption{Extrapolation curves of IP (at VBM), EA (at CBM), and indirect band gap (VBM to CBM) for \ce{BN} obtained using the ABC extrapolation model with different \kmesh ranges. All calculations are carried out using the GTH-HF pseudopotential and the GTH-cc-pVDZ basis set. The black dashed line indicates the experimental band gap that is corrected for zero-point renormalization.}
\label{fig:bn ipea}
\end{figure*}

\begin{figure*}[!htb]
\centering
\includegraphics[width=6.25in]{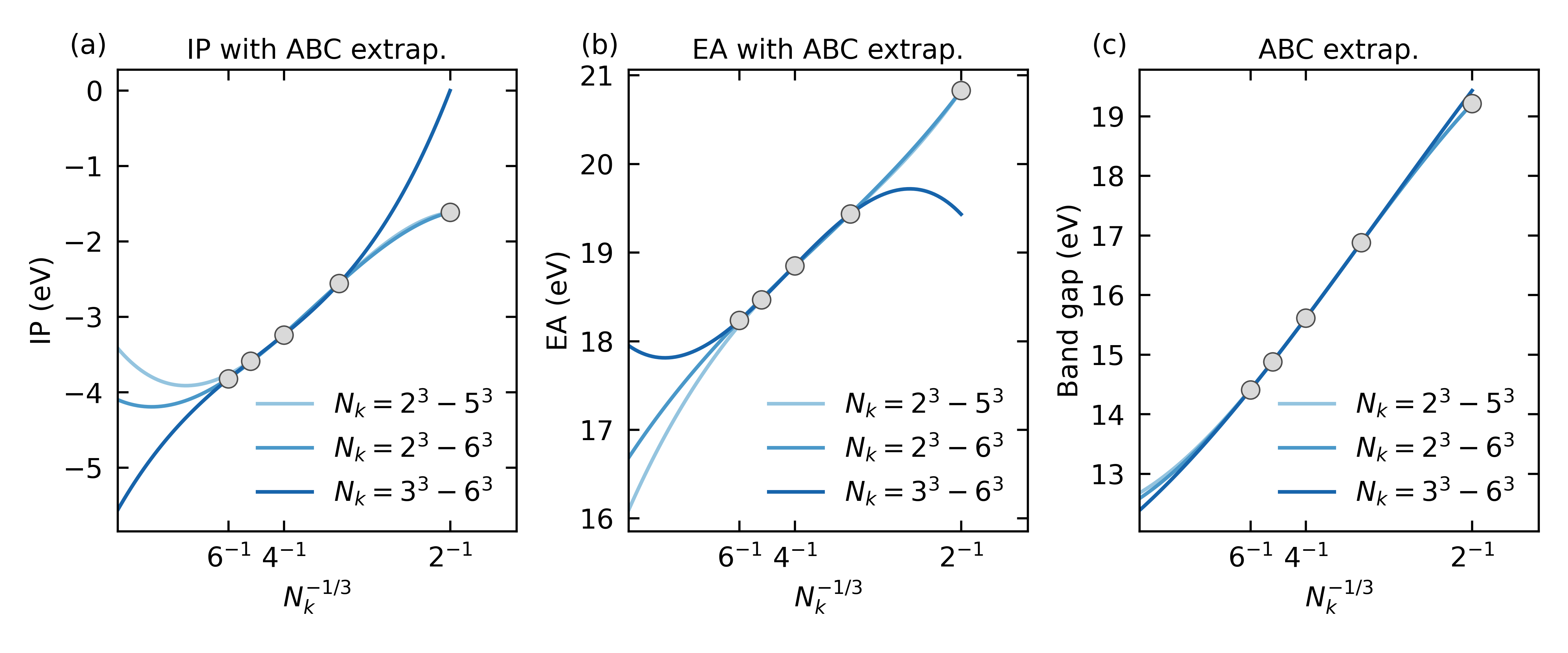} 
\caption{Extrapolation curves of IP (at CBM), EA (at CBM), and direct band gap (at CBM) for \ce{BN} obtained using the ABC extrapolation model with different \kmesh ranges. All calculations are carried out using the GTH-HF pseudopotential and the GTH-cc-pVDZ basis set.}
\label{fig:bn ipea cbm}
\end{figure*}

\begin{table*}[t]
\centering
\caption{Difference (TZ$-$DZ and QZ$-$DZ, in eV) between EOM-CCSD direct band gaps computed with the GTH-cc-pVDZ and GTH-cc-pVTZ basis sets for different systems and \kmeshes. For BN, BP, Si, C, we compute the $\Gamma \rightarrow \Gamma$ direct band gap.}
\label{tab:tz direct}
\begin{tabular}{lSSSSSSS}
\toprule
\multirow{2}{*}{System} & \multicolumn{4}{c}{TZ$-$DZ} & \multicolumn{3}{c}{QZ$-$DZ} \\
\cmidrule(lr){6-8}
\cmidrule(lr){2-5}
& \multicolumn{1}{c}{\mesh{2}}
& \multicolumn{1}{c}{\mesh{3}}
& \multicolumn{1}{c}{\mesh{4}} 
& \multicolumn{1}{c}{\mesh{5}} 
& \multicolumn{1}{c}{\mesh{2}}
& \multicolumn{1}{c}{\mesh{3}}
& \multicolumn{1}{c}{\mesh{4}} \\
\midrule
MgO  & -0.024 & 0.004 & 0.003 & 0.005 & -0.003 & 0.032 & 0.034 \\
LiCl & -0.069 & -0.050 & -0.048 & -0.045 & 0.022 & 0.049 & 0.055\\
LiF  & 0.005 & -0.007 & -0.020 & -0.026 & 0.064 & 0.055 & 0.045 \\
LiH  & -0.002 & -0.002 & 0.002 & 0.004 & -0.009 & 0.001 & 0.006\\
BN  & 0.024 & 0.042 & 0.052 & 0.057 & 0.062 & 0.083 & N/A\\
BP & -0.061 & -0.046 & -0.036 & -0.029 & -0.060 & -0.043 & N/A\\
Si  & -0.011 & -0.013 & -0.005 & 0.005 & 0.007 & 0.007 & 0.018\\
C & -0.063 & -0.041 & -0.032 & -0.027 & -0.067 & -0.041 & -0.030\\
\bottomrule
\end{tabular}
\end{table*}

\begin{table*}[t]
\centering
\caption{Conduction-band offsets (in eV) for different solids computed with different \kmeshes. All calculations are carried out using the GTH-HF pseudopotential and the GTH-cc-pVDZ basis set.}
\label{tab:cb_offset}
\begin{tabular}{lSSSSS}
\toprule
\kmesh 
& \multicolumn{1}{c}{BN}
& \multicolumn{1}{c}{BP}
& \multicolumn{1}{c}{Si}
& \multicolumn{1}{c}{C (approx.\ $k_\mathrm{CBM}$)}
& \multicolumn{1}{c}{C ($k_\mathrm{CBM}$)} \\
\midrule
\mesh{2} & -5.293 & -2.746 & -2.510 & -2.454 & -2.487 \\
\mesh{3} & -5.351 & -2.636 & -2.463 & -2.056 & -2.141 \\
\mesh{4} & -5.303 & -2.619 & -2.441 & -2.095 & -2.164 \\
\mesh{5} & -5.321 & -2.582 & -2.405 & -2.065 & -2.114 \\
\mesh{6} & -5.320 & {N/A\textsuperscript{$a$}}  & {N/A}\textsuperscript{$a$}  & -2.048 & {N/A}\textsuperscript{$a$}\\
\bottomrule
\end{tabular}\\
\textsuperscript{$a$} Conduction-band energy at $k_\mathrm{CBM}$ is not available.
\end{table*}

\begin{table*}[t]
\centering
\caption{Difference (TZ$-$DZ and QZ$-$DZ, in eV) between conduction-band offsets computed with the GTH-cc-pVDZ and GTH-cc-pVTZ basis sets for different systems and \kmeshes.}
\label{tab:tz offset}
\begin{tabular}{lSSSSS}
\toprule
\multirow{2}{*}{System} & \multicolumn{3}{c}{TZ$-$DZ} & \multicolumn{2}{c}{QZ$-$DZ} \\
\cmidrule(lr){5-6}
\cmidrule(lr){2-4}
& \multicolumn{1}{c}{\mesh{2}}
& \multicolumn{1}{c}{\mesh{3}}
& \multicolumn{1}{c}{\mesh{4}} 
& \multicolumn{1}{c}{\mesh{2}}
& \multicolumn{1}{c}{\mesh{3}} \\
\midrule
BN  & -0.022 & -0.001 & -0.006 & -0.025 & -0.003 \\
BP  & -0.038 & -0.059 & -0.042 & -0.031 & -0.054 \\
Si  & 0.006 & 0.010 & 0.025 & 0.018 & 0.021 \\
C   & 0.010 & -0.048 & -0.010 & 0.018 & -0.047 \\
\bottomrule
\end{tabular}
\end{table*}

\begin{table*}[t]
\centering
\caption{Predicted cohesive energies (in eV/atom) obtained with the GTH-HF pseudopotential (PP) and the all-electron (AE) Hamiltonian at each level of theory, compared to previous literature results.
}
\label{tab:cohesive_comp}
\begin{tabular}{lSSSSSSSSSSSS}
\toprule
{\multirow{2}{*}{System}} 
& \multicolumn{3}{c}{HF}
& \multicolumn{6}{c}{MP2}
& \multicolumn{3}{c}{CCSD} \\
\cmidrule(lr){2-4}
\cmidrule(lr){5-10}
\cmidrule(lr){11-13}
& \multicolumn{1}{c}{PP} & \multicolumn{1}{c}{AE} & \multicolumn{1}{c}{VASP~\cite{gruneis.2010.10.1063/1.3466765}} & \multicolumn{1}{c}{PP} & \multicolumn{1}{c}{AE} & \multicolumn{1}{c}{VASP~\cite{gruneis.2010.10.1063/1.3466765}} & \multicolumn{1}{c}{VASP~\cite{booth.2013.10.1038/nature11770}} & \multicolumn{1}{c}{FHI-aims~\cite{zhang.2019.10.1088/1367-2630/aaf751}} & \multicolumn{1}{c}{PySCF~\cite{goldzak.2022.10.1063/5.0119633}} & \multicolumn{1}{c}{PP} & \multicolumn{1}{c}{AE} & \multicolumn{1}{c}{VASP~\cite{booth.2013.10.1038/nature11770}} \\
\midrule
MgO	 & 3.61 & 3.63 & 3.59 & 5.47 & 5.52 & 5.35 &      & 5.53 & 5.37 & 5.09 & 5.10 &    \\
LiCl & 2.70 & 2.71 & 2.70 & 3.70 & 3.72 & 3.64 &      & 3.70 & 3.69 & 3.52 & 3.55 &    \\
LiF	 & 3.32 & 3.34 & 3.34 & 4.56 & 4.61 & 4.49 &      & 4.54 & 4.58 & 4.38 & 4.50 &    \\
LiH	 & 1.79 & 1.79 & 1.79 & 2.38 & 2.39 & 2.39 & 2.39 & 2.38 & 2.41 & 2.45 & 2.50 & 2.45   \\
BN   & 4.61 & 4.71 & 4.74 & 7.02 & 7.15 & 7.12 & 7.15 & 7.25 & 7.13 & 6.42 & 6.54 & 6.57  \\
BP   & 3.37 & 3.40 & 3.38 & 5.59 & 5.64 & 5.61 &      & 5.77 & 5.58 & 4.89 & 4.95 &    \\
Si   & 3.02 & 3.02 & 2.97 & 5.06 & 5.07 & 5.05 &      & 5.21 & 4.97 & 4.46 & 4.47 &    \\
C    & 5.13 & 5.27 & 5.28 & 7.82 & 7.99 & 7.97 & 8.04 & 8.08 & 7.98 & 7.10 & 7.26 & 7.30   \\
\bottomrule
\end{tabular}
\end{table*}

\begin{table*}[t]
\centering
\caption{Predicted EOM-CCSD band gaps (in eV) obtained with the GTH-HF pseudopotential (PP) and the all-electron (AE) Hamiltonian, compared to previous literature results. PySCF results from Ref.~\citenum{vo.2024.10.1063/5.0187856} used extrapolation model A with \meshto{3}{4}, and VASP and FHI-aims results from Ref.~\citenum{moerman.2025.10.1103/PhysRevB.111.L121202} used the GW-EOM-23 and GW-EOM-234 models, respectively.
}
\label{tab:bandgap_comp}
\begin{tabular}{lSSSSSSSSSSSS}
\toprule
System 
& \multicolumn{1}{c}{PP} & \multicolumn{1}{c}{AE} & \multicolumn{1}{c}{PySCF~\cite{vo.2024.10.1063/5.0187856}} & \multicolumn{1}{c}{VASP~\cite{moerman.2025.10.1103/PhysRevB.111.L121202}} & \multicolumn{1}{c}{FHI-aims~\cite{moerman.2025.10.1103/PhysRevB.111.L121202}} \\
\midrule
MgO  &  9.22 &  9.33 &  8.34 &  9.52 & 9.19  \\
LiCl & 10.18 & 10.36 &  9.43 &  9.90         \\
LiF  & 15.93 & 16.03 & 15.43 & 16.19        \\
LiH  &  6.28 &  6.21 &  5.85 &  6.26 & 6.32  \\
BN   &  6.62 &  6.73 &  6.45 &  6.62         \\
BP   &  2.05 &  2.07 &  1.65 &  2.27 & 2.38  \\
Si   &  1.29 &  1.23 &  0.96 &  1.29         \\
C    &  5.55 &  5.63 &  4.88 &  5.75 & 6.15 \\
\bottomrule
\end{tabular}
\end{table*}

\clearpage

\bibliography{ref}